\documentclass[a4paper, 12pt]{article}

\pdfoutput = 1

\usepackage{styleBuding}
\usepackage{bm}
\usepackage{color,listings}
\usepackage[usenames,dvipsnames,svgnames,table]{xcolor}
\usepackage{amsthm, amsmath}
\usepackage{amssymb}
\usepackage{mathrsfs}
\usepackage{graphicx}
\usepackage{slashed}
\usepackage{soul}
\usepackage{multirow}
\usepackage{subfigure}
\usepackage{diagbox}
\usepackage{siunitx} % for physics units
\usepackage{url} %for link in @article
\usepackage[utf8]{inputenc}
\usepackage[figuresright]{rotating}
\definecolor{back}{HTML}{F8F8F8}
\usepackage{epstopdf}
\usepackage{lipsum}
\newcommand\blfootnote[1]{%
	\begingroup
	\renewcommand\thefootnote{}\footnote{#1}%
	\addtocounter{footnote}{-1}%
	\endgroup
}

\newcommand{\rom}[1]{\uppercase\expandafter{\romannumeral #1\relax}}

\allowdisplaybreaks
\DeclareUnicodeCharacter{2212}{-}

\let\jnfont=\rm
\def\NPB#1,{{\jnfont Nucl.\ Phys.\ B }{\bf #1},}
\def\PLB#1,{{\jnfont Phys.\ Lett.\ B }{\bf #1},}
\def\EPJC#1,{{\jnfont Eur.\ Phys.\ Jour.\ C }{\bf #1},}
\def\PRD#1,{{\jnfont Phys.\ Rev.\ D }{\bf #1},}
\def\PRL#1,{{\jnfont Phys.\ Rev.\ Lett.\ }{\bf #1},}
\def\MPLA#1,{{\jnfont Mod.\ Phys.\ Lett.\ A }{\bf #1},}
\def\JPG#1,{{\jnfont J.\ Phys.\ G}{\bf #1},}
\def\CTP#1,{{\jnfont Commun.\ Theor.\ Phys.\ }{\bf #1},}
\def\ZPC#1,{{\jnfont Z.\ Phys.\ C }{\bf #1},}
\def\JHEP#1,{{\jnfont JHEP \ }{\bf #1},}

\title{Impact of recent measurement of $(g-2)_\mu$, LHC search for supersymmetry, and LZ experiment on Minimal Supersymmetric Standard Model}

\author{Yangle He$^{a}$, Lei Meng$^{a,*}$\blfootnote{*Corresponding author.}, Yuanfang Yue$^{a}$, and Di Zhang$^{a}$ }
\affiliation{ $^a$ Department of Physics, Henan Normal University, Xinxiang 453007, China}
\emailAdd{heyangle@htu.edu.cn}
\emailAdd{mel18@foxmail.com}
\emailAdd{yueyuanfang@htu.edu.cn}
\emailAdd{dz481655@gmail.com}

\abstract{Motivated by the recent measurement of muon anomalous magnetic moment at Fermilab, the rapid progress of the LHC search for supersymmetry, and the significantly improved sensitivities of dark matter direct detection experiments, we studied their impacts on the Minimal Supersymmetric Standard Model (MSSM). We conclude that higgsino mass should be larger than about $500~{\rm GeV}$ for $M_1 < 0 $ and $630~{\rm GeV}$ for $M_1 > 100~{\rm GeV}$, where $M_1$ denotes the bino mass. These improved bounds imply a tuning of ${\cal{O}}(1\%)$ to predict the $Z$-boson mass and simultaneously worsen the naturalness of the $Z$- and $h$-mediated resonant annihilations to achieve the measured dark matter density. We also conclude that the LHC restrictions have set lower bounds on the sparticle mass spectra: $ m_{\tilde{\chi}_1^0} \gtrsim 210~{\rm GeV}$, $m_{\tilde{\chi}_2^0}, m_{\tilde{\chi}_1^\pm} \gtrsim 235~{\rm GeV}$, $m_{\tilde{\chi}_3^0} \gtrsim 515~{\rm GeV}$, $m_{\tilde{\chi}_4^0} \gtrsim 525~{\rm GeV}$, $m_{\tilde{\chi}_2^\pm} \gtrsim 530~{\rm GeV}$, $m_{\tilde{\nu}_\mu} \gtrsim 235~{\rm GeV}$, $ m_{\tilde{\mu}_1} \gtrsim 215~{\rm GeV}$, and $m_{\tilde{\mu}_2} \gtrsim 250~{\rm GeV}$, where $\tilde{\chi}_{2}^0$ and $\tilde{\chi}_1^\pm$ are wino-dominated when they are lighter than about $500~{\rm GeV}$. These bounds are far beyond the reach of the LEP experiments in searching for supersymmetry and have not been acquired before. In addition, we illuminate how some parameter spaces of the MSSM have been tested at the LHC and provide five scenarios in which the theory coincides with the LHC restrictions. Once the muon g-2 anomaly is confirmed to originate from supersymmetry, this research may serve as a guide to explore the characteristics of the MSSM in future experiments.

}

\begin{document}
    \maketitle
    \flushbottom
\newpage

\section{Introduction}

As the cornerstone of particle physics, the Standard Model (SM) has encapsulated our best understanding of fundamental particles and forces. Although it is well tested by many experimental results, there are still unsolved puzzles, such as the quadratic divergence in the Higgs squared mass and the absence of dark matter (DM) candidates. Historically, these puzzles were viewed as the robust evidence of new physics beyond the SM, and looking for mechanisms to circumvent them was the model-building guideline. Supersymmetry (SUSY) is the most promising among the new physics theories due to its elegant structure and remarkable advantages in solving these puzzles~\cite{Fayet:1976cr,Haber:1984rc,Martin:1997ns,Jungman:1995df}.

To date, rich information about SUSY has been accumulated due to the rapid progress of particle physics experiments in recent years.
The Run-II data of the Large Hadron Collider (LHC) enabled scientists to explore the properties of winos, higgsinos, and scalar leptons (sleptons), which are the SUSY partners of $W$, Higgs, and lepton fields, respectively. It was found that wino masses up to about $1060~{\rm GeV}$ for $m_{\tilde{\chi}_1^0} \lesssim 400~{\rm GeV}$ and higgsino masses up to $900~{\rm GeV}$ for $m_{\tilde{\chi}_1^0} \lesssim 240~{\rm GeV}$ have been excluded in the simplified model of SUSY~\cite{ATLAS:2019lff}, where $\tilde{\chi}_1^0$ denotes the lightest neutralino, acting as the lightest supersymmetric particle (LSP) and thus a DM candidate under the assumption of $R$-parity conservation~\cite{Jungman:1995df}, and $m_{\tilde{\chi}_1^0}$ is its mass. The data also excluded sleptons lighter than approximately $700~{\rm GeV}$ when the LSP was massless based on statistical methods~\cite{ATLAS:2019lff,CMS:2020bfa}. Furthermore, the LUX-ZEPLIN (LZ) experiment just released its first results about the direct search for DM, where the sensitivities to spin-independent (SI) and spin-dependent (SD) cross sections of DM-nucleon scattering have reached about $6.0 \times 10^{-48}~{\rm cm^2}$ and $1.0 \times 10^{-42}~{\rm cm^2}$, respectively, for the DM mass around $30~{\rm GeV}$~\cite{LZ:2022ufs}. These unprecedented precision values strongly limit the DM coupling to the SM particles,
which are determined by SUSY parameters. In addition, the combined measurement of the muon anomalous magnetic moment, $a_{\mu} \equiv (g-2)_\mu/2$, by the E821 experiment at the Brookhaven National Laboratory (BNL)~\cite{Bennett:2006fi} and the E989 experiment at Fermilab~\cite{Abi:2021gix} indicates a 4.2$\sigma$ discrepancy from the SM's prediction~\cite{Aoyama:2020ynm,Aoyama:2012wk,Aoyama:2019ryr,Czarnecki:2002nt,Gnendiger:2013pva,Davier:2017zfy,Keshavarzi:2018mgv,Colangelo:2018mtw,Hoferichter:2019gzf,Davier:2019can,Keshavarzi:2019abf,Kurz:2014wya,Melnikov:2003xd,Masjuan:2017tvw,Colangelo:2017fiz,Hoferichter:2018kwz,Gerardin:2019vio,Bijnens:2019ghy,Colangelo:2019uex,Blum:2019ugy,Colangelo:2014qya}. Although this difference may have been induced by the uncertainties in calculating the hadronic contribution to the moment, as revealed by the recent lattice simulation of the BMW collaboration~\cite{Borsanyi:2020mff}, it was widely speculated to arise from new physics (see, e.g., Ref.~\cite{Athron:2021iuf} and the references therein). Along this direction, it is remarkable that once the difference is confirmed to originate from SUSY effects, salient features of the theory, e.g., the mass spectra of the electroweakinos and sleptons, can be inferred ~\cite{Martin:2001st,Domingo:2008bb,Moroi:1995yh,Hollik:1997vb,Athron:2015rva,Endo:2021zal,Stockinger:2006zn,Czarnecki:2001pv,Cao:2011sn,Endo:2013lva,Kang:2016iok,Zhu:2016ncq,Yanagida:2017dao, Hagiwara:2017lse,Cox:2018qyi,Tran:2018kxv,Padley:2015uma,
	Choudhury:2017fuu,Okada:2016wlm,Du:2017str, Ning:2017dng, Wang:2018vxp,Yang:2018guw,Liu:2020nsm,Cao:2019evo,Cao:2021lmj,Ke:2021kgy,Lamborn:2021snt,Li:2021xmw,Nakai:2021mha,Li:2021koa,Kim:2021suj,Li:2021pnt,Altmannshofer:2021hfu,
	Baer:2021aax,Chakraborti:2021bmv,Aboubrahim:2021xfi,Iwamoto:2021aaf,Chakraborti:2021dli,Cao:2021tuh,Yin:2021mls,Zhang:2021gun,Ibe:2021cvf,Zheng:2021wnu,
	Han:2021ify,Wang:2021bcx,Zheng:2021gug,Chakraborti:2021mbr,Aboubrahim:2021myl,Ali:2021kxa,Wang:2021lwi,Chakraborti:2020vjp,Baum:2021qzx,Gu:2021mjd,Cao:2022chy,Cao:2022htd,Domingo:2022pde,Cao:2022ovk,Wang:2023suf,Zhao:2022pnv,Yang:2022qyz,Sabatta:2019nfg,Cao:2023juc}.

Given that SUSY predictions on these experimental results rely on different theoretical inputs, it is essential to collectively study their impacts on the Minimal Supersymmetric Standard Model (MSSM), which is the most economical realization of SUSY in particle physics~\cite{Haber:1984rc,Gunion:1984yn,Djouadi:2005gj}. For this purpose, we organize this study as follows. In Sec. \ref{theory-section}, we briefly introduce the basics of the MSSM, including its DM physics, the SUSY contribution to $a_\mu$, the signals of SUSY particles (sparticles) at the LHC, and the strategy to search for them.
In Sec. \ref{numerical study}, we perform a sophisticated scan over the broad parameter space of the MSSM and clarify how the MSSM remains consistent with the experimental results. Finally, we draw conclusions in Sec. \ref{conclusion-section}.

%---------------------------------------------
\section{\label{theory-section} Theoretical preliminaries of MSSM}

The following superpotential of the MSSM was given in Ref.~\cite{Haber:1984rc,Gunion:1984yn}:
\begin{eqnarray}
    W = - Y_d  \hat{q} \cdot \hat{H}_d \ \hat{d} - Y_e \hat{l} \cdot \hat{H}_d\ \hat{e}
    	+ Y_u \hat{q} \cdot \hat{H}_u\ \hat{u} + \mu \hat{H}_u \cdot \hat{H}_d,
    \label{MSSMSF}
\end{eqnarray}
where the superfields $\hat{q}$ and $\hat{l}$ are left-handed SU(2) doublets for quarks and leptons, respectively, and $\hat{u}$, $\hat{d}$, and $\hat{e}$ are right-handed singlets for the fermions. The scalar components of the Higgs doublet superfields, $\hat{H}_u$ and $\hat{H}_d$, are given by $H_u=(H_u^+,H_u^0)$ and $H_d=(H_d^0,H_d^-)$, respectively, and their product is defined by $H_u \cdot H_d = (H_u^+ H_d^- -H_u^0 H_d^0)$. The first three terms in the superpotential represent the Yukawa couplings of the quark and lepton fields, and the last term is responsible for the higgsino mass.

The MSSM predicts two CP-even Higgs bosons, $h$ and $H$, one CP-odd Higgs boson $A$, and a pair of charged Higgs boson $H^\pm = \cos \beta H_u^\pm + \sin \beta H_d^\pm$ in the Higgs sector~\cite{Gunion:1984yn,Djouadi:2005gj}. Among these states, $h$ denotes the SM-like scalar discovered at the LHC with $m_h \simeq 125~{\rm GeV}$, and the neutral states $H$ and $A$ are approximately degenerate with $H^\pm$ in mass.
The LHC search for non-SM-like Higgs bosons has obtained model-independent upper limits on the production rates of $H$, $A$, and $H^\pm$ (see, e.g., Ref.~\cite{ATLAS:2020zms,ATLAS:2021upq}), indicating that they should be massive. The electroweakino sector of the MSSM consists of four neutralinos and two pairs of charginos~\cite{Gunion:1984yn}, denoted by $\tilde{\chi}_i^0$ with $i=1,2,3,4$ and $\tilde{\chi}_j^\pm$ with $j=1,2$, respectively, in this work. The neutralinos are superpositions of bino ($\tilde{B}$), wino ($\tilde{W}^0$), and two higgsino fields ($\tilde{H}_d^0$ and $\tilde{H}_u^0$), and they are majorana fermions. By contrast, the left-handed and right-handed components of the chargino $\tilde{\chi}^+_j$ come from the mixing of $\tilde{W}^+$ with $\tilde{H}_u^+$ and $\tilde{W}^-$ with $\tilde{H}_d^-$ respectively, and $\tilde{\chi}_j^\pm$ are Dirac fermions. By convention, the neutralinos as mass eigenstates are labeled in an ascending mass order, and so are the charginos. In addition, each slepton mass eigenstate in the MSSM is associated with a definite flavor quantum number if there is no flavor mixing in the slepton sector~\cite{Dimopoulos:1995ju,Gabbiani:1996hi}. The $\ell$-flavored sleptons $\tilde{\ell}_i$ (i=1,2) are mixtures of chiral scalar fields $\tilde{\ell}_L$ and $\tilde{\ell}_R$. Given that the mixing is usually small, we also denote $\tilde{\ell}_i$ by its dominant component sometimes to facilitate our discussion.

%------------------------------------------
\subsection{DM physics in MSSM }

On the premise of explaining both the measured DM density and the muon g-2 anomaly, the DM candidate in the MSSM must be the bino-dominated lightest neutralino~\cite{Bagnaschi:2017tru}\footnote{In the case that the lightest left-handed sneutrino acts as a DM candidate, its interaction with the $Z$-boson predicts a much smaller density than its measured value, i.e., $\Omega h^2 \ll 0.12$, and meanwhile an unacceptably large DM-nucleon scattering rate~\cite{Falk:1994es}. For the wino- or higgsino-dominated DM case, the density is below $10^{-3}$ by our calculation. These cases were surveyed in the MSSM to explain the muon g-2 anomaly in Ref.~\cite{Chakraborti:2021kkr}.}.
It achieves the measured density through the co-annihilation with wino-dominated electroweakinos or sleptons, or through the $Z$- or $h$-mediated resonant annihilation~\cite{Griest:1988ma}. In the co-annihilation case, the reactions $S_i S_j \to X X^\prime$, where $S_i S_j$ may be any of LSP-LSP, LSP-NLSP (next-to-lightest supersymmetric particle), and NLSP-NLSP annihilation states and $X X^\prime$ denotes SM particles, contribute to the density~\cite{Griest:1990kh,Baker:2015qna}. The effective annihilation rate at a temperature $T$ is then given by Eq. (3.2) in Ref.~\cite{Baker:2015qna}. This formula indicates that the annihilation partner has a significant effect only when the departure of its mass from the DM mass is less than about $10\%$. The resonant annihilation is distinct in that the density is very sensitive to the splitting between $2 |m_{\tilde{\chi}_1^0}|$ and the mediator's mass~\cite{Griest:1988ma}. The weaker the DM coupled to the mediator, the smaller the splitting must be to achieve the measured density. Evidently, this situation requires the fine-tuning of the theoretical parameters.

The cross sections of the DM-nucleon scattering take the following form~\cite{Baum:2017enm,Cao:2019qng,Cao:2021ljw}
\begin{eqnarray}
\sigma_{\tilde{\chi}_1^0-N}^{\rm SI} & & \simeq  5 \times 10^{-45} {\rm cm^2} \times \left ( \frac{F_u^N + F_d^N}{0.28} \right )^2 \times \left \{ \frac{F_u^N}{F_u^N + F_d^N} \times \right . \nonumber \\
& &  \left [  \frac{\cos \alpha}{\sin \beta} \left ( \frac{C_{\tilde{\chi}_1^0 \tilde{\chi}_1^0 h}}{0.1} \right ) \left ( \frac{125~{\rm GeV}}{m_h} \right )^2 +  \frac{\sin \alpha}{\sin \beta} \left ( \frac{C_{\tilde{\chi}_1^0 \tilde{\chi}_1^0 H}}{0.1} \right ) \left ( \frac{125~{\rm GeV}}{m_H} \right )^2 \right ] +  \frac{F_d^N}{F_u^N + F_d^N}   \nonumber \\
& & \left . \times \left [  - \frac{\sin \alpha}{\cos \beta} \left ( \frac{C_{\tilde{\chi}_1^0 \tilde{\chi}_1^0 h}}{0.1} \right ) \left ( \frac{125~{\rm GeV}}{m_h} \right )^2 +  \frac{\cos \alpha}{\cos \beta} \left ( \frac{C_{\tilde{\chi}_1^0 \tilde{\chi}_1^0 H}}{0.1} \right ) \left ( \frac{125~{\rm GeV}}{m_H} \right )^2 \right ] \right \}^2, \label{SI-1}  \\
\sigma_{\tilde{\chi}_1^0-N}^{\rm SD} && \simeq  C_N \times \left ( \frac{C_{\tilde{\chi}_1^0 \tilde{\chi}_1^0 Z}}{0.1} \right )^2,
\end{eqnarray}
where $F_u^N$ and $F_d^N$ denote the normalized up-type and down-type quark contributions to the nucleon mass, respectively, and $C_N$ is related with the nucleon spin with $C_p \simeq 1.8 \times 10^{-40}~{\rm cm^2} $ for protons and $C_n \simeq 1.4 \times 10^{-40}~{\rm cm^2} $ for neutrons. $\alpha$ is the mixing angle of the CP-even Higgs states satisfying $\alpha \simeq \beta - \pi/2$ in the large $m_A$ limit~\cite{Djouadi:2005gj}, and $\beta$ is defined by the ratio of the Higgs vacuum expectation values, namely $\tan \beta \equiv v_u/v_d$. $C_{\tilde{\chi}_1^0 \tilde{\chi}_1^0 h}$, $C_{\tilde{\chi}_1^0 \tilde{\chi}_1^0 H}$, and $C_{\tilde{\chi}_1^0 \tilde{\chi}_1^0 Z}$ represent the DM couplings to the Higgs bosons $h$ and $H$, and $Z$-boson, respectively. In the series expansion with $m_Z/\mu$ as a variable, they are approximated by~\cite{Pierce:2013rda,Huang:2014xua,Calibbi:2014lga,Cheung:2014lqa}:
\begin{eqnarray}
C_{\tilde{\chi}_1^0 \tilde{\chi}_1^0 h} & \simeq & e \tan \theta_W \frac{m_Z}{\mu (1 - m_{\tilde{\chi}_1^0}^2/\mu^2)} \left ( \cos (\beta + \alpha) + \sin (\beta - \alpha) \frac{m_{\tilde{\chi}_1^0}}{\mu} \right )  \nonumber  \\
& \simeq & e \tan \theta_W \frac{m_Z}{\mu (1 -  m_{\tilde{\chi}_1^0}^2/\mu^2)} \left ( \sin 2 \beta + \frac{m_{\tilde{\chi}_1^0}}{\mu} \right ), \label{SI-2} \\
C_{\tilde{\chi}_1^0 \tilde{\chi}_1^0 H} & \simeq & e \tan \theta_W \frac{m_Z}{\mu (1 - m_{\tilde{\chi}_1^0}^2/\mu^2)} \left ( \sin (\beta + \alpha) + \cos (\beta - \alpha) \frac{m_{\tilde{\chi}_1^0}}{\mu} \right )  \nonumber  \\
& \simeq & - e \tan \theta_W \cos 2 \beta \frac{m_Z}{\mu (1 -  m_{\tilde{\chi}_1^0}^2/\mu^2)}, \label{SI-3} \\
C_{\tilde{\chi}_1^0 \tilde{\chi}_1^0 Z} & \simeq & \frac{e \tan \theta_W \cos 2 \beta}{2} \frac{m_Z^2}{\mu^2 - m_{\tilde{\chi}_1^0}^2},  \label{SD-2}
\end{eqnarray}
where $\theta_W$ is the weak mixing angle, and the DM mass $ m_{\tilde{\chi}_1^0}$ relates to the bino mass $M_1$ by $ m_{\tilde{\chi}_1^0} \simeq M_1$. Taking $F_u^N \simeq F_d^N \simeq 0.14$~\cite{Huang:2014xua} and $\tan \beta \gg 1$, one can conclude that
\begin{eqnarray}
\sigma_{\tilde{\chi}_1^0-N}^{\rm SI} & \simeq &  1.4 \times 10^{-44} {\rm cm^2} \times \left ( \frac{m_Z}{\mu (1- m_{\tilde{\chi}_1^0}^2/\mu^2)} \right )^2  \nonumber \\
 & & \times \left [ (\sin 2 \beta + \frac{m_{\tilde{\chi}_1^0}}{\mu} ) \left (\frac{125~{\rm GeV}}{m_h} \right )^2  - \frac{\tan \beta}{2} \left (\frac{125~{\rm GeV}}{m_H} \right )^2 \right ]^2.  \label{Final-SI-Approximation}
\end{eqnarray}
This formula reveals that if $M_1$ and $\mu$ are of the same sign and the Higgs boson $H$ is tremendously massive, $\mu$ must be sufficiently large to be consistent with the results of the PandaX-4T experiment~\cite{PandaX-4T:2021bab}. It also shows that if $M_1$ and $\mu$ are of opposite signs, which can result in the blind spots of the scattering~\cite{Huang:2014xua,Crivellin:2015bva,Han:2016qtc,Carena:2018nlf}, $|\mu| \sim 100~{\rm GeV}$ seems to be experimentally allowed. However, such a possibility has been limited by the experimental search for the SD DM-nucleon scattering because, regardless of the relative sign between $M_1$ and $\mu$, a small $|\mu|$ can enhance the scattering cross section. In summary, the DM direct detection experiments alone have set a lower bound on the magnitude of $\mu$. With the improvement of the experimental sensitivity, the bound will become tightened.

\subsection{\label{DMRD}Muon g-2 }

The SUSY source of the muon g-2, $a^{\rm SUSY}_{\mu}$, mainly includes loops mediated by a smuon and a neutralino and those containing a muon-flavor sneutrino and a chargino~\cite{Moroi:1995yh,Domingo:2008bb,Hollik:1997vb,Martin:2001st}. The full one-loop contributions to $a^{\rm SUSY}_{\mu}$ in the MSSM are not presented here for brevity. Instead, we provide the expression of $a_\mu^{\rm SUSY}$ in the mass insertion approximation to reveal its key features~\cite{Moroi:1995yh}. Specifically, at the lowest order of the approximation, the contributions to $a_\mu^{\rm SUSY}$ are divided into four types: WHL, BHL, BHR, and BLR, where $W$, $B$, $H$, $L$, and $R$ represent wino, bino, higgsino, and left-handed and right-handed smuon fields, respectively. They arise from the Feynman diagrams involving $\tilde{W}-\tilde{H}_d$, $\tilde{B}-\tilde{H}_d^0$, $\tilde{B}-\tilde{H}_d^0$, and $\tilde{\mu}_L-\tilde{\mu}_R$ transitions, respectively, and take the following forms~\cite{Athron:2015rva, Moroi:1995yh,Endo:2021zal}:
\begin{eqnarray}
a_{\mu, \rm WHL}^{\rm SUSY}
    &=&\frac{\alpha_2}{8 \pi} \frac{m_{\mu}^2 M_2 \mu \tan \beta}{m_{\tilde{\nu}_\mu}^4} \left \{ 2 f_C\left(\frac{M_2^2}{m_{\tilde{\nu}_{\mu}}^2}, \frac{\mu^2}{m_{\tilde{\nu}_{\mu}}^2} \right) - \frac{m_{\tilde{\nu}_\mu}^4}{\tilde{m}_{\tilde{\mu}_L}^4} f_N\left(\frac{M_2^2}{\tilde{m}_{\tilde{\mu}_L}^2}, \frac{\mu^2}{\tilde{m}_{\tilde{\mu}_L}^2} \right) \right \}\,, \quad \quad
    \label{eq:WHL} \\
a_{\mu, \rm BHL}^{\rm SUSY}
  &=& \frac{\alpha_Y}{8 \pi} \frac{m_\mu^2 M_1 \mu  \tan \beta}{\tilde{m}_{\tilde{\mu}_L}^4} f_N\left(\frac{M_1^2}{\tilde{m}_{\tilde{\mu}_L}^2}, \frac{\mu^2}{\tilde{m}_{\tilde{\mu}_L}^2} \right)\,,
    \label{eq:BHL} \\
a_{\mu, \rm BHR}^{\rm SUSY}
  &=& - \frac{\alpha_Y}{4\pi} \frac{m_{\mu}^2 M_1 \mu \tan \beta}{\tilde{m}_{\tilde{\mu}_R}^4} f_N\left(\frac{M_1^2}{\tilde{m}_{\tilde{\mu}_R}^2}, \frac{\mu^2}{\tilde{m}_{\tilde{\mu}_R}^2} \right)\,,
    \label{eq:BHR} \\
a_{\mu, \rm BLR}^{\rm SUSY}
  &=& \frac{\alpha_Y}{4\pi} \frac{m_{\mu}^2  M_1 \mu \tan \beta}{M_1^4}
    f_N\left(\frac{\tilde{m}_{\tilde{\mu}_L}^2}{M_1^2}, \frac{\tilde{m}_{\tilde{\mu}_R}^2}{M_1^2} \right)\,,
    \label{eq:BLR}
\end{eqnarray}
where $\tilde{m}_{\tilde{\mu}_L}$ and $\tilde{m}_{\tilde{\mu}_R}$ are soft-breaking masses for left-handed and right-handed smuon fields, respectively, at the slepton mass scale, and they are approximately equal to slepton masses.  The loop functions are given by
\begin{eqnarray}
    \label{eq:loop-aprox}
    f_C(x,y)
    &=&  \frac{5-3(x+y)+xy}{(x-1)^2(y-1)^2} - \frac{2\ln x}{(x-y)(x-1)^3}+\frac{2\ln y}{(x-y)(y-1)^3} \,,
      \\
    f_N(x,y)
    &=&
      \frac{-3+x+y+xy}{(x-1)^2(y-1)^2} + \frac{2x\ln x}{(x-y)(x-1)^3}-\frac{2y\ln y}{(x-y)(y-1)^3} \,,
\end{eqnarray}
satisfying $f_C(1,1) = 1/2$ and $f_N(1,1) = 1/6$.

The following points about $a_\mu^{\rm SUSY}$ should be noted:
\begin{itemize}
\item If all the dimensional SUSY parameters involved in $a_\mu^{\rm SUSY}$ take a common value $M_{\rm SUSY}$, $a_\mu^{\rm SUSY}$ is proportional to $m_\mu^2 \tan \beta/M_{\rm SUSY}^2$, indicating that the muon g-2 anomaly prefers a large $\tan \beta$ and a moderately low SUSY scale.
\item Unlike the "WHL", "BHL", and "BHR" contributions, which usually diminish monotonously with the increase of $|\mu|$, the "BLR" contribution is linearly proportional to $\mu$. As a result, the tremendously massive higgsino scenario, characterized by predicting $\mu \gtrsim 30~{\rm TeV}$, can yield the central value of the muon g-2 anomaly even when
    $M_1$, $M_{\tilde{\mu}_L}$, and $M_{\tilde{\mu}_R}$ are at the TeV scale~\cite{Gu:2021mjd}. In this study, we are not interested in this case since it needs severe fine tunings to predict $m_Z$~\cite{Baer:2012uy}.
\item Assuming $|\mu| < 1~{\rm TeV}$, the "WHL" contribution is usually much larger than the other contributions if $\tilde{\mu}_L$ is not significantly heavier than $\tilde{\mu}_R$~\cite{Cao:2021tuh}.
\item The difference between the $a_\mu^{\rm SUSY}$ values calculated by the mass insertion approximation and the full expression is less than $3\%$. We verified this conclusion for the green samples in Fig.~\ref{fig1} of this work.
\item The two-loop (2L) contributions to $a_\mu$, including 2L corrections to SM one-loop diagrams and those to SUSY one-loop diagrams~\cite{Stockinger:2006zn}, are about $-5\%$
of the one-loop prediction~\cite{Cao:2022ovk}. These were neglected in this study.
\end{itemize}

\begin{table}[]
	\caption{Experimental analyses of the electroweakino production processes considered in this study, which are categorized by the topologies of the supersymmetry (SUSY) signals.}
	\label{Table1}
	\vspace{0.2cm}
	\resizebox{0.98\textwidth}{!}{
		\begin{tabular}{llll}
			\hline\hline
			\texttt{Scenario} & \texttt{Final State} &\multicolumn{1}{c}{\texttt{Name}}\\\hline
			\multirow{6}{*}{$\tilde{\chi}_{2}^0\tilde{\chi}_1^{\pm}\rightarrow WZ\tilde{\chi}_1^0\tilde{\chi}_1^0$}&\multirow{6}{*}{$n\ell (n\geq2) + nj(n\geq0) + \text{E}_\text{T}^{\text{miss}}$}&\texttt{CMS-SUS-20-001($137fb^{-1}$)}~\cite{CMS:2020bfa}\\&&\texttt{ATLAS-2106-01676($139fb^{-1}$)}~\cite{ATLAS:2021moa}\\&&\texttt{CMS-SUS-17-004($35.9fb^{-1}$)}~\cite{CMS:2018szt}\\&&\texttt{CMS-SUS-16-039($35.9fb^{-1}$)}~\cite{CMS:2017moi}\\&&\texttt{ATLAS-1803-02762($36.1fb^{-1}$)}~\cite{ATLAS:2018ojr}\\&&\texttt{ATLAS-1806-02293($36.1fb^{-1}$)}~\cite{ATLAS:2018eui}\\\\
			\multirow{2}{*}{$\tilde{\chi}_2^0\tilde{\chi}_1^{\pm}\rightarrow \ell\tilde{\nu}\ell\tilde{\ell}$}&\multirow{2}{*}{$n\ell (n=3) + \text{E}_\text{T}^{\text{miss}}$}&\texttt{CMS-SUS-16-039($35.9fb^{-1}$)}~\cite{CMS:2017moi}\\&&\texttt{ATLAS-1803-02762($36.1fb^{-1}$)}~\cite{ATLAS:2018ojr}\\\\
			$\tilde{\chi}_2^0\tilde{\chi}_1^{\pm}\rightarrow \tilde{\tau}\nu\ell\tilde{\ell}$&$2\ell + 1\tau + \text{E}_\text{T}^{\text{miss}}$&\texttt{CMS-SUS-16-039($35.9fb^{-1}$)}~\cite{CMS:2017moi}\\\\
			$\tilde{\chi}_2^0\tilde{\chi}_1^{\pm}\rightarrow \tilde{\tau}\nu\tilde{\tau}\tau$&$3\tau + \text{E}_\text{T}^{\text{miss}}$&\texttt{CMS-SUS-16-039($35.9fb^{-1}$)}~\cite{CMS:2017moi}\\\\
			\multirow{6}{*}{$\tilde{\chi}_{2}^0\tilde{\chi}_1^{\pm}\rightarrow Wh\tilde{\chi}_1^0\tilde{\chi}_1^0$}&\multirow{6}{*}{$n\ell(n\geq1) + nb(n\geq0) + nj(n\geq0) + \text{E}_\text{T}^{\text{miss}}$}&\texttt{ATLAS-1909-09226($139fb^{-1}$)}~\cite{ATLAS:2020pgy}\\&&\texttt{CMS-SUS-17-004($35.9fb^{-1}$)}~\cite{CMS:2018szt}\\&&\texttt{CMS-SUS-16-039($35.9fb^{-1}$)}~\cite{CMS:2017moi}\\
			&&\texttt{ATLAS-1812-09432($36.1fb^{-1}$)}\cite{ATLAS:2018qmw}\\&&\texttt{CMS-SUS-16-034($35.9fb^{-1}$)}\cite{CMS:2017kxn}\\&&\texttt{CMS-SUS-16-045($35.9fb^{-1}$)}~\cite{CMS:2017bki}\\\\
			\multirow{2}{*}{$\tilde{\chi}_1^{\mp}\tilde{\chi}_1^{\pm}\rightarrow WW\tilde{\chi}_1^0 \tilde{\chi}_1^0$}&\multirow{2}{*}{$2\ell + \text{E}_\text{T}^{\text{miss}}$}&\texttt{ATLAS-1908-08215($139fb^{-1}$)}~\cite{ATLAS:2019lff}\\&&\texttt{CMS-SUS-17-010($35.9fb^{-1}$)}~\cite{CMS:2018xqw}\\\\
			\multirow{2}{*}{$\tilde{\chi}_1^{\mp}\tilde{\chi}_1^{\pm}\rightarrow 2\tilde{\ell}\nu(\tilde{\nu}\ell)$}&\multirow{2}{*}{$2\ell + \text{E}_\text{T}^{\text{miss}}$}&\texttt{ATLAS-1908-08215($139fb^{-1}$)}~\cite{ATLAS:2019lff}\\&&\texttt{CMS-SUS-17-010($35.9fb^{-1}$)}~\cite{CMS:2018xqw}\\\\
			$\tilde{\chi}_2^{0}\tilde{\chi}_1^{\mp}\rightarrow h/ZW\tilde{\chi}_1^0\tilde{\chi}_1^0,\tilde{\chi}_1^0\rightarrow \gamma/Z\tilde{G}$&\multirow{2}{*}{$2\gamma + n\ell(n\geq0) + nb(n\geq0) + nj(n\geq0) + \text{E}_\text{T}^{\text{miss}}$}&\multirow{2}{*}{\texttt{ATLAS-1802-03158($36.1fb^{-1}$)}~\cite{ATLAS:2018nud}}\\$\tilde{\chi}_1^{\pm}\tilde{\chi}_1^{\mp}\rightarrow WW\tilde{\chi}_1^0\tilde{\chi}_1^0,\tilde{\chi}_1^0\rightarrow \gamma/Z\tilde{G}$&&\\\\
			$\tilde{\chi}_2^{0}\tilde{\chi}_1^{\pm}\rightarrow ZW\tilde{\chi}_1^0\tilde{\chi}_1^0,\tilde{\chi}_1^0\rightarrow h/Z\tilde{G}$&\multirow{4}{*}{$n\ell(n\geq4) + \text{E}_\text{T}^{\text{miss}}$}&\multirow{4}{*}{\texttt{ATLAS-2103-11684($139fb^{-1}$)}~\cite{ATLAS:2021yyr}}\\$\tilde{\chi}_1^{\pm}\tilde{\chi}_1^{\mp}\rightarrow WW\tilde{\chi}_1^0\tilde{\chi}_1^0,\tilde{\chi}_1^0\rightarrow h/Z\tilde{G}$&&\\$\tilde{\chi}_2^{0}\tilde{\chi}_1^{0}\rightarrow Z\tilde{\chi}_1^0\tilde{\chi}_1^0,\tilde{\chi}_1^0\rightarrow h/Z\tilde{G}$&&\\$\tilde{\chi}_1^{\mp}\tilde{\chi}_1^{0}\rightarrow W\tilde{\chi}_1^0\tilde{\chi}_1^0,\tilde{\chi}_1^0\rightarrow h/Z\tilde{G}$&&\\\\
			\multirow{3}{*}{$\tilde{\chi}_{i}^{0,\pm}\tilde{\chi}_{j}^{0,\mp}\rightarrow \tilde{\chi}_1^0\tilde{\chi}_1^0+\chi_{soft}\rightarrow ZZ/H\tilde{G}\tilde{G}$}&\multirow{3}{*}{$n\ell(n\geq2) + nb(n\geq0) + nj(n\geq0) + \text{E}_\text{T}^{\text{miss}}$}&\texttt{CMS-SUS-16-039($35.9fb^{-1}$)}~\cite{CMS:2017moi}\\&&\texttt{CMS-SUS-17-004($35.9fb^{-1}$)}~\cite{CMS:2018szt}\\&&\texttt{CMS-SUS-20-001($137fb^{-1}$)}~\cite{CMS:2020bfa}\\\\
			\multirow{2}{*}{$\tilde{\chi}_{i}^{0,\pm}\tilde{\chi}_{j}^{0,\mp}\rightarrow \tilde{\chi}_1^0\tilde{\chi}_1^0+\chi_{soft}\rightarrow HH\tilde{G}\tilde{G}$}&\multirow{2}{*}{$n\ell(n\geq2) + nb(n\geq0) + nj(n\geq0) + \text{E}_\text{T}^{\text{miss}}$}&\texttt{CMS-SUS-16-039($35.9fb^{-1}$)}~\cite{CMS:2017moi}\\&&\texttt{CMS-SUS-17-004($35.9fb^{-1}$)}~\cite{CMS:2018szt}\\\\
			$\tilde{\chi}_{2}^{0}\tilde{\chi}_{1}^{\pm}\rightarrow W^{*}Z^{*}\tilde{\chi}_1^0\tilde{\chi}_1^0$&$3\ell + \text{E}_\text{T}^{\text{miss}}$&\texttt{ATLAS-2106-01676($139fb^{-1}$)}~\cite{ATLAS:2021moa}\\\\
			\multirow{3}{*}{$\tilde{\chi}_{2}^{0}\tilde{\chi}_{1}^{\pm}\rightarrow Z^{*}W^{*}\tilde{\chi}_1^0\tilde{\chi}_1^0$}&\multirow{2}{*}{$2\ell + nj(n\geq0) + \text{E}_\text{T}^{\text{miss}}$}&\texttt{ATLAS-1911-12606($139fb^{-1}$)}~\cite{ATLAS:2019lng}\\&&\texttt{ATLAS-1712-08119($36.1fb^{-1}$)}~\cite{ATLAS:2017vat}\\&&\texttt{CMS-SUS-16-048($35.9fb^{-1}$)}~\cite{CMS:2018kag}\\\\
			\multirow{3}{*}{$\tilde{\chi}_{2}^{0}\tilde{\chi}_{1}^{\pm}+\tilde{\chi}_{1}^{\pm}\tilde{\chi}_{1}^{\mp}+\tilde{\chi}_{1}^{\pm}\tilde{\chi}_{1}^{0}$}&\multirow{3}{*}{$2\ell + nj(n\geq0) + \text{E}_\text{T}^{\text{miss}}$}&\texttt{ATLAS-1911-12606($139fb^{-1}$)}~\cite{ATLAS:2019lng}\\&&\texttt{ATLAS-1712-08119($36.1fb^{-1}$)}~\cite{ATLAS:2017vat}\\&&\texttt{CMS-SUS-16-048($35.9fb^{-1}$)}~\cite{CMS:2018kag}\\\hline
					
	\end{tabular}} % }
\end{table}

\begin{table}[]
	\caption{Same as Table~\ref{Table1}, but for the slepton production processes.}
	\label{Table2}
  \centering
	\vspace{0.2cm}
	\resizebox{0.7\textwidth}{!}{
		\begin{tabular}{llll}
			\hline\hline
			\texttt{Scenario} & \texttt{Final State} &\multicolumn{1}{c}{\texttt{Name}}\\\hline
\multirow{6}{*}{$\tilde{\ell}\tilde{\ell}\rightarrow \ell\ell\tilde{\chi}_1^0\tilde{\chi}_1^0$}&\multirow{6}{*}{$2\ell + \text{E}_\text{T}^{\text{miss}}$}&\multirow{1}{*}{\texttt{ATLAS-1911-12606($139fb^{-1}$)}~\cite{ATLAS:2019lng}}\\&&\multirow{1}{*}{\texttt{ATLAS-1712-08119($36.1fb^{-1}$)}~\cite{ATLAS:2017vat}}\\&&\multirow{1}{*}{\texttt{ATLAS-1908-08215($139fb^{-1}$)}~\cite{ATLAS:2019lff}}\\&&\multirow{1}{*}{\texttt{CMS-SUS-20-001($137fb^{-1}$)}~\cite{CMS:2020bfa}}\\&&\multirow{1}{*}{\texttt{ATLAS-1803-02762($36.1fb^{-1}$)}~\cite{ATLAS:2018ojr}}\\&&\multirow{1}{*}{\texttt{CMS-SUS-17-009($35.9fb^{-1}$)}~\cite{CMS:2018eqb}}\\\hline

\end{tabular}} % }
\end{table}

\vspace{-0.3cm}

\subsection{LHC search for SUSY}

Since some of the electroweakinos and sleptons involved in $a_\mu^{\rm SUSY}$ must be moderately light to account for the anomaly~\cite{Chakraborti:2020vjp}, they are copiously produced at the LHC and thus are subjected to strong constraints from the SUSY searches at the LHC with $\sqrt{s}=13~\rm{TeV}$. These searches usually concentrate on theories with $R$-parity conservation~\cite{Fayet:1977yc,Farrar:1978xj}, where the LSP is undetected, leading to missing energy in the final states. We implement these restrictions by
scrutinizing the experimental analyses in Tables~\ref{Table1} and~\ref{Table2}. We find that the following reports are particularly critical:
\begin{itemize}
\item \texttt{CMS-SUS-20-001~\cite{CMS:2020bfa}}: Search for SUSY signal containing two oppositely charged same-flavor leptons and missing transverse momentum. This analysis studied not only strong sparticle productions but also electroweakino productions. The lepton originated from an on-shell or off-shell $Z$ boson in the decay chain or from the decay of the produced sleptons. For the electroweakino pair production, the wino-dominated chargino and neutralino were explored up to masses of $750~{\rm GeV}$ and $800~{\rm GeV}$, respectively. For the slepton pair production, the first two-generation sleptons were explored up to a mass of $700~{\rm GeV}$.

\item \texttt{CMS-SUS-16-039 and CMS-SUS-17-004~\cite{CMS:2017moi,CMS:2018szt}}: Search for electroweakino productions with two, three, or four leptons and missing transverse momentum ($\rm{E}_{\rm{T}}^{\rm{miss}}$) in the final states. One remarkable strategy of this analysis was that it included all the possible final states and defined several categories by the number of leptons in the event, their flavors, and their charges to enhance the discovery potential. In the context of simplified models, the observed limit on wino-dominated $m_{\tilde{\chi}_1^{\pm}}$ in the chargino--neutralino production was about 650 GeV for the $WZ$ topology, 480 GeV for the $WH$ topology, and 535 GeV for the mixed topology.

\item \texttt{ATLAS-2106-01676~\cite{ATLAS:2021moa}}: Search for wino- or higgsino-dominated chargino--neutralino pair productions. This analysis investigated on-shell $WZ$, off-shell $WZ$, and $Wh$ categories in the decay chain and focused on the final state containing exactly three leptons, possible ISR jets, and $\rm{E}_{\rm{T}}^{\rm{miss}}$.
    For the wino scenario in the simplified model, the exclusion bound of $m_{\tilde{\chi}_2^0}$ was about $640~\rm{GeV}$ for a massless $\tilde{\chi}_1^0$, and it was weakened
    as the mass difference between $\tilde{\chi}_2^0$ and $\tilde{\chi}_1^0$ diminished. Specifically, $\tilde{\chi}_2^0$ should be heavier than about $500~\rm{GeV}$ for $m_{\tilde{\chi}_1^0} = 300~{\rm GeV}$ (the on-shell $W/Z$ case), $300~\rm{GeV}$ for a positive $m_{\tilde{\chi}_1^0}$ and $ 35~{\rm GeV} \lesssim m_{\tilde{\chi}_2^0} - m_{\tilde{\chi}_1^0} \lesssim 90~{\rm GeV}$ (the off-shell $W/Z$ case), and $220~\rm{GeV}$ when $ m_{\tilde{\chi}_2^0} - m_{\tilde{\chi}_1^0} = 15~{\rm GeV}$ (the extreme off-shell $W/Z$ case). By contrast, $\tilde{\chi}_2^0$ was excluded only up to a mass of $210~\rm{ GeV}$ for the off-shell $W/Z$ case of the higgsino scenario, which occurred when
    $ m_{\tilde{\chi}_2^0} - m_{\tilde{\chi}_1^0} = 10~{\rm GeV}$ or  $ m_{\tilde{\chi}_2^0} - m_{\tilde{\chi}_1^0} \gtrsim 35~{\rm GeV}$.

\item \texttt{ATLAS-1911-12606~\cite{ATLAS:2019lng}}: Concentration on compressed mass spectra case and search for electroweakino pair or slepton pair production, with two leptons and missing transverse momentum as the final state. The results were projected onto the $\Delta m-\tilde{\chi}_2^0$ plane, where $\Delta m \equiv m_{\tilde{\chi}_2^0} -  m_{\tilde{\chi}_1^0}$ for the electroweakino production. It was found that the tightest bound on the higgsino-dominated $\tilde{\chi}_2^0$ was $193~{\rm GeV}$ in mass for $\Delta m \simeq 9.3~{\rm GeV}$, and the optimum bound on the wino-dominated $\tilde{\chi}_2^0$ was $240~{\rm GeV}$ in mass when $\Delta m \simeq 7~{\rm GeV}$. Similarly, it was found that light-flavor sleptons should be heavier than about 250 GeV for $\Delta m_{\tilde{\ell}} = 10~{\rm GeV}$, where $m_{\tilde{\ell}} \equiv m_{\tilde{\ell}} - m_{\tilde{\chi}_1^0}$.

\end{itemize}

Note that all the analyses were based on $139~\rm fb^{-1}$ data except for the second analysis, which studied $36~\rm fb^{-1}$ data.

\begin{table}[tbp]
\caption{Parameter space explored in this study, where $\tan \beta$ was defined at the electroweak scale and the others were defined at the renormalization scale $Q=1~{\rm TeV}$. $M_{\tilde{\mu}_L}$ and $M_{\tilde{\mu}_R}$ are soft-breaking masses for left-handed and right-handed smuon fields, respectively.  Other dimensional parameters not crucial to this study were fixed at 3 TeV, including the SUSY parameters for the first- and third-generation sleptons, three generation squarks (except for the soft trilinear coefficients $A_t$ and $A_b$, which are assumed to be equal and change freely), and gluinos. \label{Table3}}
\centering

\vspace{0.3cm}

\resizebox{0.8\textwidth}{!}{
\begin{tabular}{c|c|c||c|c|c}
\hline
Parameter & Prior & Range & Parameter & Prior & Range   \\
\hline
$\tan{\beta}$ & Flat & $1 \sim 60$ & $A_t/{\rm TeV}$ & Flat & $-5.0\sim 5.0$ \\
$\mu/{\rm TeV}$ & Log & $0.1\sim 1.0$ &$m_{\rm A}/{\rm TeV}$ & Log & $0.5\sim 10$ \\
$M_1/{\rm TeV}$ & Flat & $-1.0\sim1.0$ & $M_2/{\rm TeV}$ & Log & $0.1\sim 1.5$ \\
$M_{\tilde{\mu}_L}/{\rm TeV}$ & Log & $0.1\sim 1.0$ &$M_{\tilde{\mu}_R}/{\rm TeV}$ & Log & $0.1\sim 1.0$ \\
\hline
\end{tabular}}
\end{table}

\section{\label{numerical study}Combined experimental impacts on MSSM}

This research utilized the package \textsf{SARAH\,4.14.3}~\cite{Staub:2008uz, Staub:2012pb, Staub:2013tta, Staub:2015kfa} to build the model file of the MSSM,
the codes \textsf{SPheno\,4.0.4}~\cite{Porod:2003um, Porod:2011nf} and \textsf{FlavorKit}~\cite{Porod:2014xia} to generate particle mass spectra and compute low energy
observables, such as $a_\mu^{\rm SUSY}$ and $B$-physics observables,  and the package \textsf{MicrOMEGAs\,5.0.4}~\cite{Belanger:2001fz, Belanger:2005kh, Belanger:2006is, Belanger:2010pz, Belanger:2013oya, Barducci:2016pcb} to calculate DM observables, assuming that the lightest neutralino was the sole DM candidate in the universe. Bounds from the direct search for extra Higgs bosons at the LEP, Tevatron, and LHC and the fit of $h$'s property to LHC Higgs data were implemented by the programs~\textsf{HiggsBounds\,5.10.2}~\cite{HB2008jh,HB2011sb,HB2013wla,HB2020pkv} and \textsf{HiggsSignal\,2.6.2}~\cite{HS2013xfa,HSConstraining2013hwa,HS2014ewa,HS2020uwn}, respectively.

%\vspace{-0.2cm}

\subsection{\label{scan}Research strategy}

The main aims of this research were to explore as many possibilities (parameter points) of the MSSM as possible, clarify how they remain consistent with current experimental results, and reveal some distinct characteristics of the theory. We carried out such a study by the following procedures:
\begin{itemize}
\item We employed the \textsf{MultiNest} algorithm~\cite{Feroz:2008xx} to comprehensively scan the parameter space in Table~\ref{Table3}. The $n_{\rm live}$ parameter in the algorithm controlled the number of active points sampled in each iteration of the scan, and $n_{\rm live} = 10000$ was set. The following likelihood function was constructed to guide the scan:
\begin{eqnarray}
\mathcal{L} = \mathcal{L}_{a_\mu} \times \mathcal{L}_{const}. \label{likelihood}
\end{eqnarray}
$\mathcal{L}_{a_\mu}$ is the likelihood function of the muon $g-2$ anomaly given by
\begin{eqnarray}
\mathcal{L}_{a_\mu} \equiv Exp\left[-\frac{1}{2} \left( \frac{a_{\mu}^{\rm SUSY}- \Delta a_\mu}{\delta a_\mu }\right)^2\right] = Exp\left[-\frac{1}{2} \left( \frac{a_{\mu}^{\rm SUSY}- 2.51\times 10^{-9}}{5.9\times 10^{-10} }\right)^2\right], \nonumber
\end{eqnarray}
where $\Delta a_\mu \equiv a_\mu^{\rm Exp} - a_\mu^{\rm SM}$ and $\delta a_\mu$ represent the difference between the experimental central value of $a_\mu$ and its SM prediction and the total uncertainties in determining $\Delta a_\mu$, respectively~\cite{Abi:2021gix,Bennett:2006fi,Aoyama:2020ynm,Aoyama:2012wk,Aoyama:2019ryr,Czarnecki:2002nt,Gnendiger:2013pva,Davier:2017zfy,
Keshavarzi:2018mgv,Colangelo:2018mtw,Hoferichter:2019gzf,Davier:2019can,Keshavarzi:2019abf,Kurz:2014wya,Melnikov:2003xd,Masjuan:2017tvw,
Colangelo:2017fiz,Hoferichter:2018kwz,Gerardin:2019vio,Bijnens:2019ghy,Colangelo:2019uex,Blum:2019ugy,Colangelo:2014qya}. $\mathcal{L}_{const}$ denotes the restrictions of some experiments on the theory. They included the consistency of $h$'s properties with the LHC Higgs data at the $95\%$ confidence level (C.L.)~\cite{HS2020uwn}, the collider searches for extra Higgs bosons~\cite{HB2020pkv}, the central value of the DM relic density from the Planck-2018 data~\cite{Planck:2018vyg} (assuming theoretical uncertainties of $20\%$ in the density calculation), the $90\%$ C.L. upper bounds of the PandaX-4T experiment on the SI DM-nucleon scattering~\cite{PandaX-4T:2021bab} and the XENON-1T experiment on the SD scattering~\cite{Aprile:2019dbj},  the $2\sigma$ bounds on the branching ratios of $B \to X_s \gamma$ and $B_s \to \mu^+ \mu^-$~\cite{PhysRevD.98.030001}, and the vacuum stability of the scalar potential consisting of the Higgs fields and the last two generations of slepton fields~\cite{Camargo-Molina:2013qva,Camargo-Molina:2014pwa}. We defined $\mathcal{L}_{const} = 1$ if the restrictions were satisfied and $\mathcal{L}_{const} = Exp[-100]$ if they were not. More details of these restrictions were introduced in Refs.~\cite{Cao:2021tuh,Cao:2022chy}.

\item We refined the samples obtained in the scan by the criteria $\mathcal{L}_{const} = 1$ and $|a_{\mu}^{\rm SUSY}- \Delta a_\mu|/\delta a_\mu \leq 3$, and we
projected those passing the selection onto the two-dimensional planes spanned by any two of the parameters $M_1$, $M_2$, $\mu$, $\tilde{m}_{\tilde{\mu}_L}$, and $\tilde{m}_{\tilde{\mu}_R}$. We then concentrated on the region of the planes where the samples were sparsely distributed and performed a new scan by adjusting relevant parameter ranges and setting $n_{\rm live} = 3000$.

\item We iterated the last operation with all accumulated samples until the projected areas on the planes remained unchanged. At this point, we acquired $2.21 \times 10^5$ samples surviving the criteria, and about $1.7 \times 10^5$ of them could further explain the $(g-2)_\mu$ anomaly at the $2\sigma$ level.

\item We simplified the study of the restrictions from the LHC search for SUSY. Specifically, given that the sample number was huge and the Monte Carlo simulation of each sample introduced below would cost more than one core-hour for our computing cluster, we selected some representative points and carried out the simulations. These points were stored in a specially designed sample database, achieved in the following way:
    \begin{enumerate}
    \item We searched for the sample with the greatest likelihood value from a database storing all the scan results.
    \item We constructed a hypersphere in the parameter space, which was centered around the sample with a radius of $10~{\rm GeV}$ for $M_1$, $M_2$, and $\mu$ (note that the simulation results were more sensitive to these three parameters than the other dimensional parameters), $20~{\rm GeV}$ for $\tilde{m}_{\tilde{\mu}_L}$ and $\tilde{m}_{\tilde{\mu}_R}$, and one unit for $\tan \beta$.
    \item We copied useful information of the central sample to the newly built database for simulation and sequentially deleted all samples in the hypersphere to update the initial database.
    \item We iterated the above operations until all the samples in the initial database were depleted.
    \end{enumerate}
We add that the method to dilute the dense samples was plausible because the $R$-value of the simulation relied heavily on both the mass spectra and the field compositions of sparticles, which determined their production cross sections and decay branching ratios. All samples in the hypersphere had similar properties in these two aspects, and the $R$-value of the studied point was typical for these samples. In addition, the method paid more attention to the samples favored by the anomaly, which was the focus of this research. After requiring the representative points to explain the muon g-2 anomaly at the $2 \sigma$ level,
we finally acquired 58242 samples for the simulations.

\item We surveyed the LHC restrictions on the representative points by simulating the following processes:
\begin{eqnarray}
pp &\to& \tilde{\chi}_i^0\tilde{\chi}_j^{\pm}, \quad i = 2, 3, 4, 5, \quad j = 1, 2; \\
pp &\to& \tilde{\chi}_i^{\pm}\tilde{\chi}_j^{\mp}, \quad i,j = 1, 2; \\
pp &\to& \tilde{\chi}_i^{0}\tilde{\chi}_j^{0}, \quad i,j = 2, 3, 4, 5; \\
pp &\to& \tilde{\mu}_i^\ast \tilde{\mu}_j,\quad i,j = 1, 2; \\
pp &\to& \tilde{\nu}_{\mu}^\ast \tilde{\nu}_\mu.
\end{eqnarray}
Specifically, the cross sections of these processes at $\sqrt{s}$ = 13 TeV were calculated at the next-to-leading order (NLO) by the package \textsf{Prospino2}~\cite{Beenakker:1996ed}, and 60000 and 40000 events were generated for the electroweakino and slepton production processes, respectively, by the package \textsf{MadGraph\_aMC@NLO}~\cite{Alwall:2011uj, Conte:2012fm}. A relevant parton shower and hadronization were completed by the program \textsf{PYTHIA8}~\cite{Sjostrand:2014zea}. The resulting event files were then fed into the package \textsf{CheckMATE\,2.0.29}~\cite{Drees:2013wra,Dercks:2016npn, Kim:2015wza} to calculate the $R$-value defined by $R \equiv max\{S_i/S_{i,obs}^{95}\}$, where $S_i$ denotes the simulated event number of the $i$-th SR in the analyses of Tables~\ref{Table1} and~\ref{Table2}, and $S_{i,obs}^{95}$ represents its corresponding $95\%$ confidence level upper limit. In this process, program \textsf{Delphes} was encoded in \textsf{CheckMATE} for detector simulation~\cite{deFavereau:2013fsa}.

As an alternative, we also used the program \textsf{SModelS\,2.2.1}~\cite{Alguero:2021dig} to study the LHC restrictions. We found that this program's capability to exclude SUSY points was usually weaker than that of the simulation due to its limited database and strict working prerequisites.

\end{itemize}

We clarified the reasons for studying the parameter space defined in Table~\ref{Table3}:
\begin{itemize}
\item Given that the soft-breaking masses of the three-generation squarks were fixed at $3~{\rm TeV}$, $A_t$ ranging from $-5~{\rm TeV}$ to $5~{\rm TeV}$ could provide an appropriate correction to acquire $ 122~{\rm GeV} \lesssim m_h \lesssim 128~{\rm GeV}$ by stop-mediated loops~\cite{Djouadi:2005gj}.
\item Since $m_A$ could significantly affect the Higgs properties and the DM-nucleon scatterings, we let it vary from $0.5~{\rm TeV}$ to $10~{\rm TeV}$, where the lower bound was inspired by the results of the LHC search for extra Higgs bosons~\cite{ATLAS:2020zms,ATLAS:2021upq} and the upper bound was chosen to be tremendously large to show the decoupling feature of the heavy Higgs bosons~\cite{Djouadi:2005gj}.
\item Motivated by the results of the LEP search for charginos\footnote{The website http://lepsusy.web.cern.ch/lepsusy/Welcome.html provided the LEP results in SUSY search. They were acquired by the LEP SUSY Working Group, consisting of ALEPH, DELPHI, L3, and OPAL collaborations.} and the naturalness to predict $Z$-boson mass~\cite{Baer:2012uy}, we assumed $0.1~{\rm TeV} \leq \mu \leq 1~{\rm TeV}$.
\item Noting that the muon g-2 anomaly preferred $|m_{\tilde{\chi}_1^0}| \lesssim 650~{\rm GeV}$~\cite{Chakraborti:2020vjp}, we set $|M_1| < 1~{\rm TeV}$.
\item $M_2$, $M_{\tilde{\mu}_L}$, and  $M_{\tilde{\mu}_R}$ in the considered ranges could explain the muon g-2 anomaly at the $2\sigma$ level and simultaneously predict the measured DM relic abundance. Their lower bounds arose from the exclusion capability of the LEP experiment in searching for SUSY, and their upper bounds were motivated by our previous interpretations of the muon g-2 anomaly in the Next-to-Minimal Supersymmetric Standard Model~\cite{Cao:2022htd}, which shared many features with the MSSM. If any of these parameters lay beyond the upper bounds, MSSM would become challenging to explain the muon g-2 anomaly on the premise of coinciding with the experimental results.
\item  We adopted the same range of $\tan \beta$ as those in the latest global fits of the MSSM to various available experimental data, which the Mastercode group performed~\cite{deVries:2015hva,Bagnaschi:2017tru,Costa:2017gup}. The range was also consistent with the most recent explanations of the muon g-2 anomaly in the MSSM (see, e.g., Refs.~\cite{Chakraborti:2020vjp,Endo:2021zal,Cox:2021nbo,Wang:2021bcx}). We noted that a large $\tan \beta$ could enhance the bottom Yukawa coupling, and the perturbativity of the theory  up to the grand unification scale implied $\tan \beta \lesssim 75$~\cite{Altmannshofer:2010zt}. For the case of $60 \leq \tan \beta \leq 75$, unexplored in this study, the electroweakinos and smuons were preferred to be heavier than those in the present work explaining the muon g-2 anomaly. Consequently, the LHC restrictions on the theory became weak, and the conclusions of this study remained unchanged. One could acquire these conclusions from the feature of $a_\mu^{\rm SUSY} \varpropto \tan \beta$ and the upper left panel of Fig.~\ref{fig2} in this work.
\end{itemize}

\begin{figure}[t]
	\centering
	\includegraphics[width=0.45\textwidth]{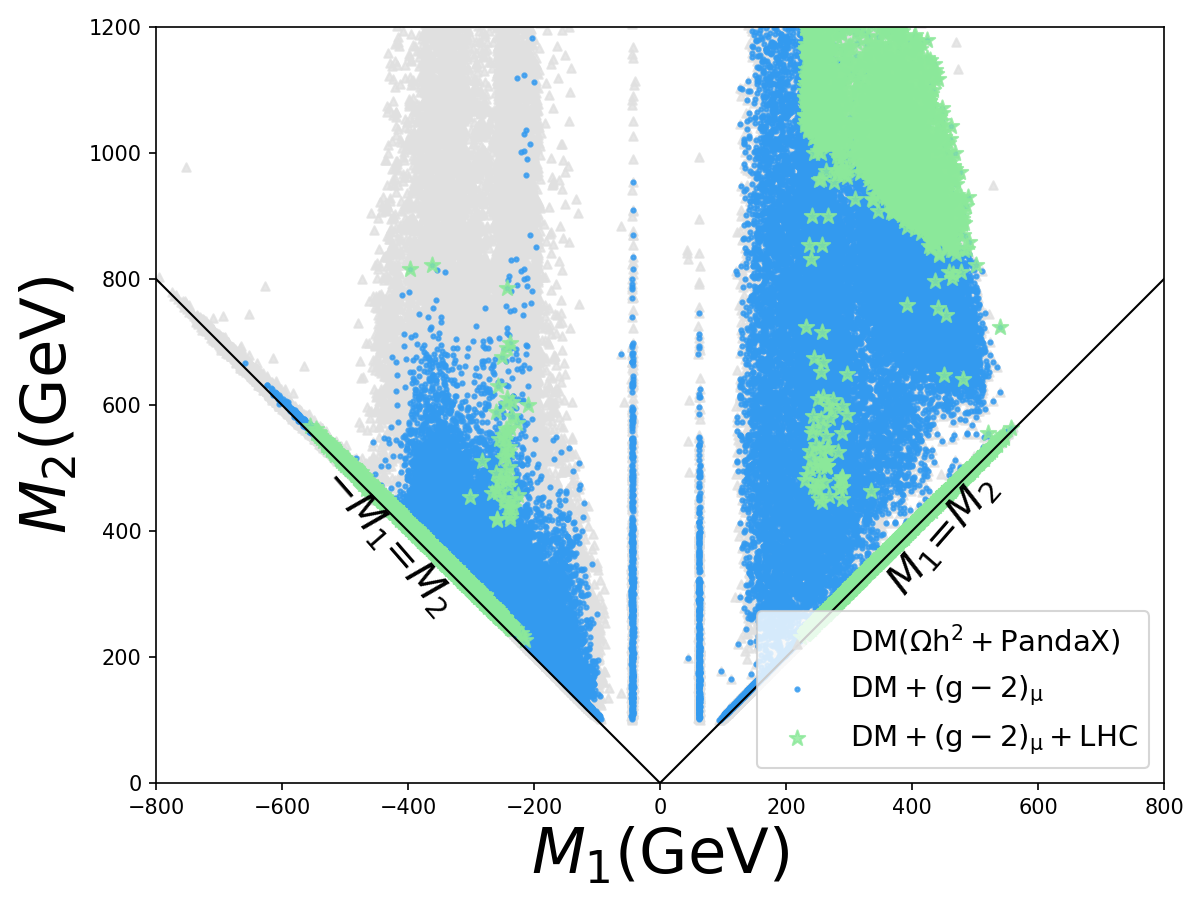}%\hspace{-0.1cm}
	\includegraphics[width=0.45\textwidth]{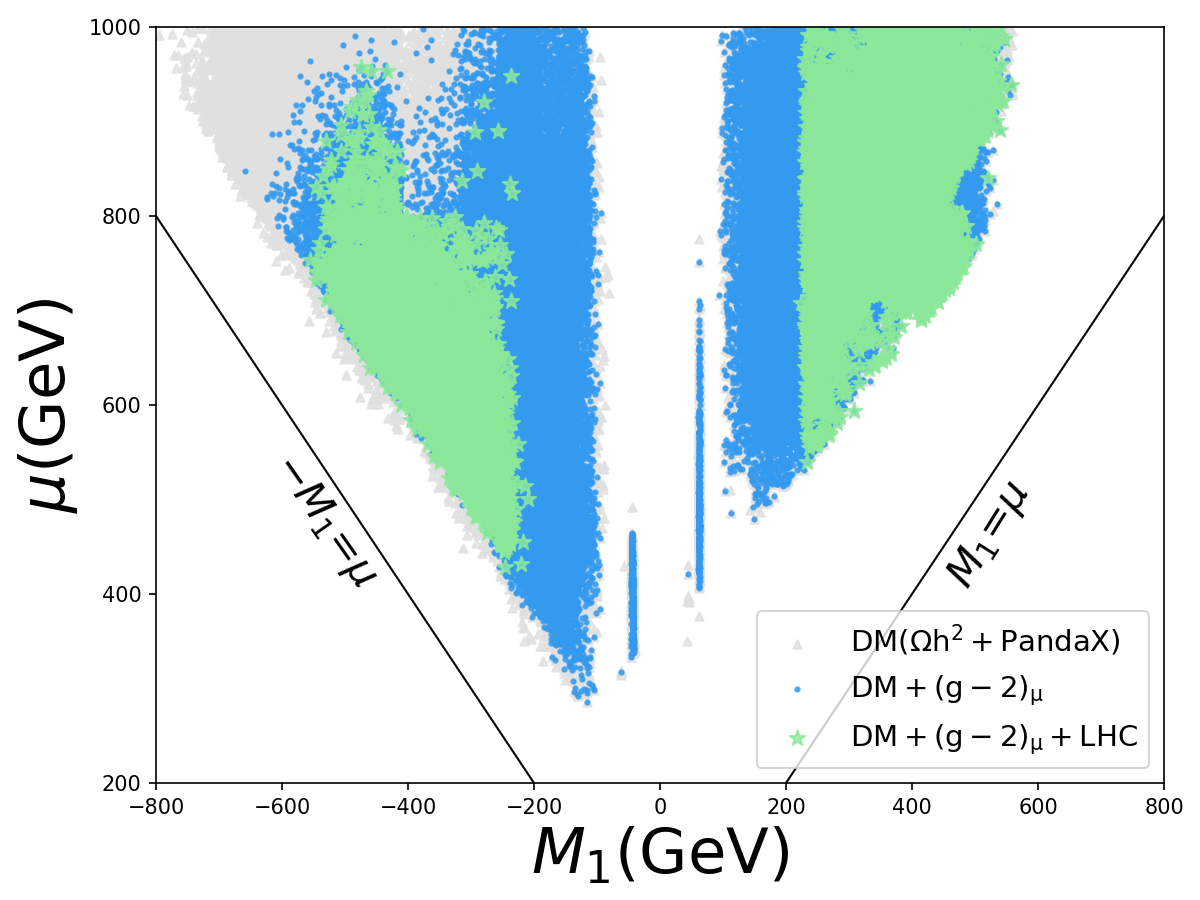}%\hspace{-0.3cm}

\caption{\label{fig1} Projection of the obtained samples onto $M_1-M_2$ plane~(left panel) and $M_1-\mu$ plane~(right panel). The gray triangles denote the samples that satisfy all restrictions listed in the text, in particular those from the DM relic density measured by the Planck experiment~\cite{Planck:2018vyg} and the DM direct detection of the PandaX-4T experiment~\cite{PandaX-4T:2021bab}. The blue circles represent those that could further explain the muon g-2 anomaly at the $2\sigma$ level, and the green stars are the part of the blue circles that agree with the results from the LHC search for SUSY. Although only about one-third of the blue circles were studied by the simulations in this work (see the research strategy in the last sub-section), the shape of the green areas is unlikely to significantly change. We verified this point by randomly selecting several thousands of samples from the blue areas and simulating the LHC restrictions.  }
\end{figure}

\subsection{\label{region}Key features of the results}

First, we studied the DM physics of the MSSM by projecting the samples acquired by the scans onto the $M_1-M_2$ and $M_1-\mu$ planes to obtain Fig.~\ref{fig1}. This figure reveals the following facts:
\begin{itemize}
\item If only the restrictions from the DM physics are considered, the DM candidate is bino-dominated for $|M_1| \leq 800~{\rm GeV}$. It may achieve the measured relic density by the $Z$-mediated resonant annihilation, the $h$-mediated resonant annihilation, or the co-annihilation with wino-like electroweakinos. In addition, we will show later in Figs.~\ref{fig5} and \ref{fig6} that it may also acquire the density by co-annihilating with $\tilde{\mu}_L/\tilde{\nu}_\mu$ or $\tilde{\mu}_R$.
\item If the samples are further required to explain the muon g-2 anomaly at the $2 \sigma$ level, $|M_1|$ is upper bounded by about $620~{\rm GeV}$. Furthermore, if the LHC restrictions are included, it is upper bounded by about $570~{\rm GeV}$.
\item There is a vacant region on the $M_1-M_2$ plane, located in the ranges of $350~{\rm GeV} \lesssim M_1 \lesssim 570~{\rm GeV}$, $ 400~{\rm GeV} \lesssim M_2 \lesssim 600~{\rm GeV}$, and $ M_1 + 30~{\rm GeV} \lesssim M_2 \lesssim M_1 + 100~{\rm GeV}$. This region is distinct because winos can significantly affect the mass of $\tilde{\nu}_\mu$ by radiative corrections, and given the values of $M_1$ and $M_2$, one needs to fine-tune the soft-breaking parameter $M_{\tilde{\mu}_L}$ to predict the measured DM density by co-annihilating with $\tilde{\nu}_\mu/\tilde{\mu}_L$. This situation is challenging in the scans since it easily results in $\tilde{\nu}_\mu$ as the LSP.

    We add that this vacant region corresponds to the void on the bottom right corner of the $|m_{\tilde{\chi}_1^0}|-m_{\tilde{\chi}_1^\pm}$ plane in Fig.~\ref{fig4}.

\item The higgsino mass $\mu$ should be larger than $300~{\rm GeV}$, $350~{\rm GeV}$, $410~{\rm GeV}$, and $500~{\rm GeV}$ for the cases of $M_1 \lesssim - 100~{\rm GeV}$, $M_1 \simeq - m_Z/2$, $M_1 \simeq m_Z/2$, and $M_1 \gtrsim 100~{\rm GeV}$, respectively. In particular, $\mu$ is more tightly limited for the possibility of $M_1 > 0$ than for the case of $M_1 < 0$, given that $|M_1|$ is fixed. These phenomena arise from the restrictions of the PandaX-4T experiment. One can understand them by the expression of $\sigma_{\tilde{\chi}_1^0-N}^{\rm SI} $ in Eq.~\ref{Final-SI-Approximation}, noting the approximation $m_{\tilde{\chi}_1^0} \simeq M_1 $ and the fact that a negative $M_1$ can lead to the cancelation of different contributions to the SI DM-nucleon scattering cross-section.

\item The observables in the DM physics and the muon g-2 anomaly may prefer different parameter spaces of the MSSM, even though broad parameter regions can still accommodate both. For example, a negative $M_1$ is disfavored by the muon g-2 anomaly in the large $\mu$ and $M_2$ region, because the BLR contribution to $a_\mu^{\rm SUSY}$ is negatively sizable. This case, however, can easily reproduce the results of DM experiments.

\end{itemize}

\begin{figure}[t]
	\centering
	\includegraphics[width=0.45\textwidth]{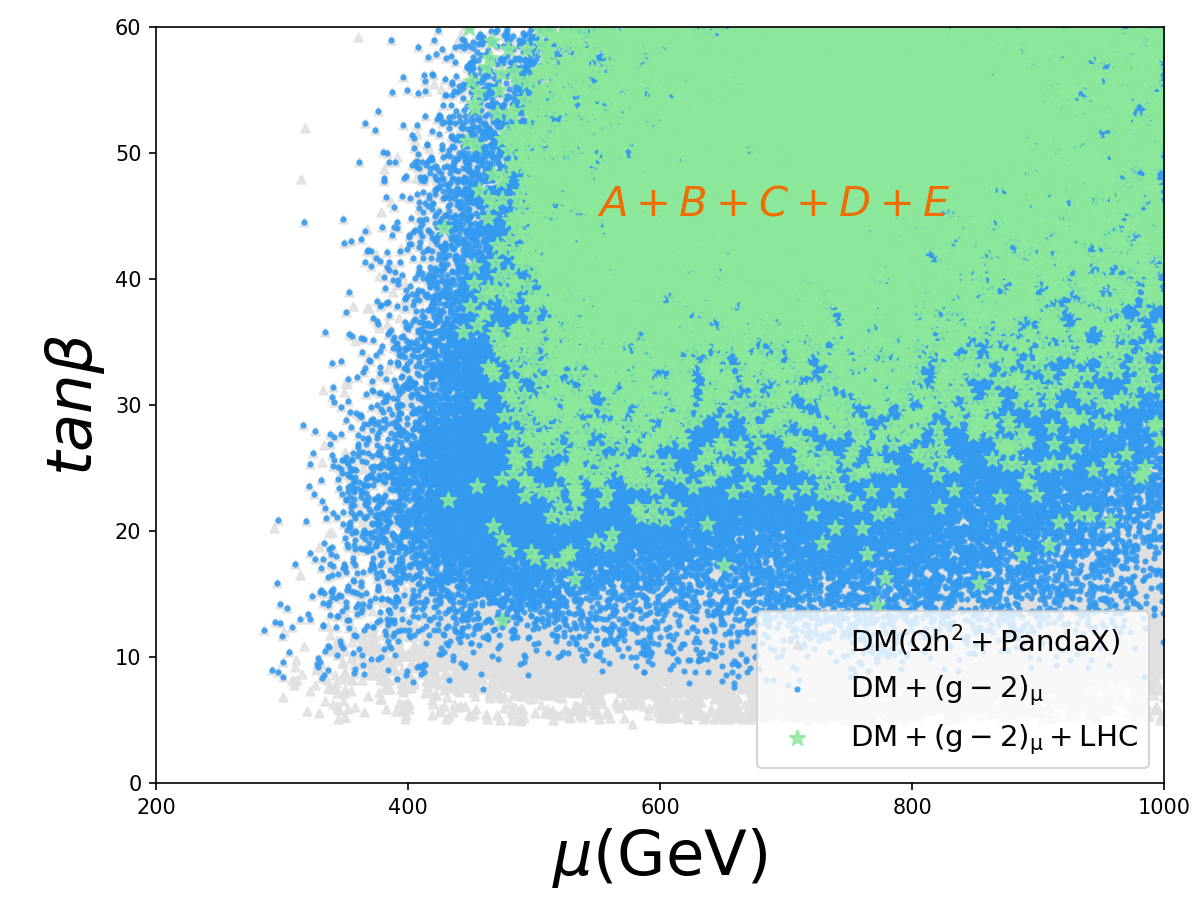}%\hspace{-0.3cm}
	\includegraphics[width=0.45\textwidth]{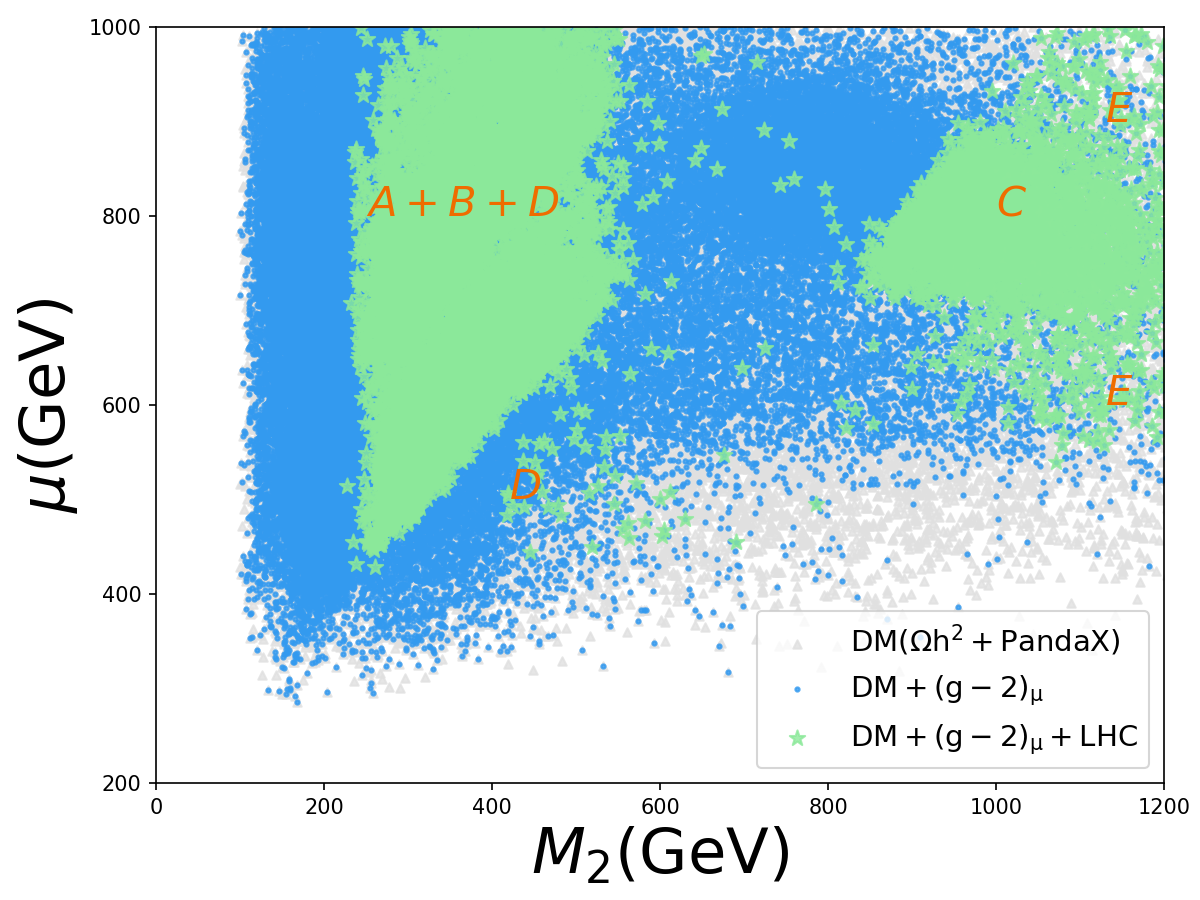}%\hspace{-0.3cm}
\\
	\includegraphics[width=0.45\textwidth]{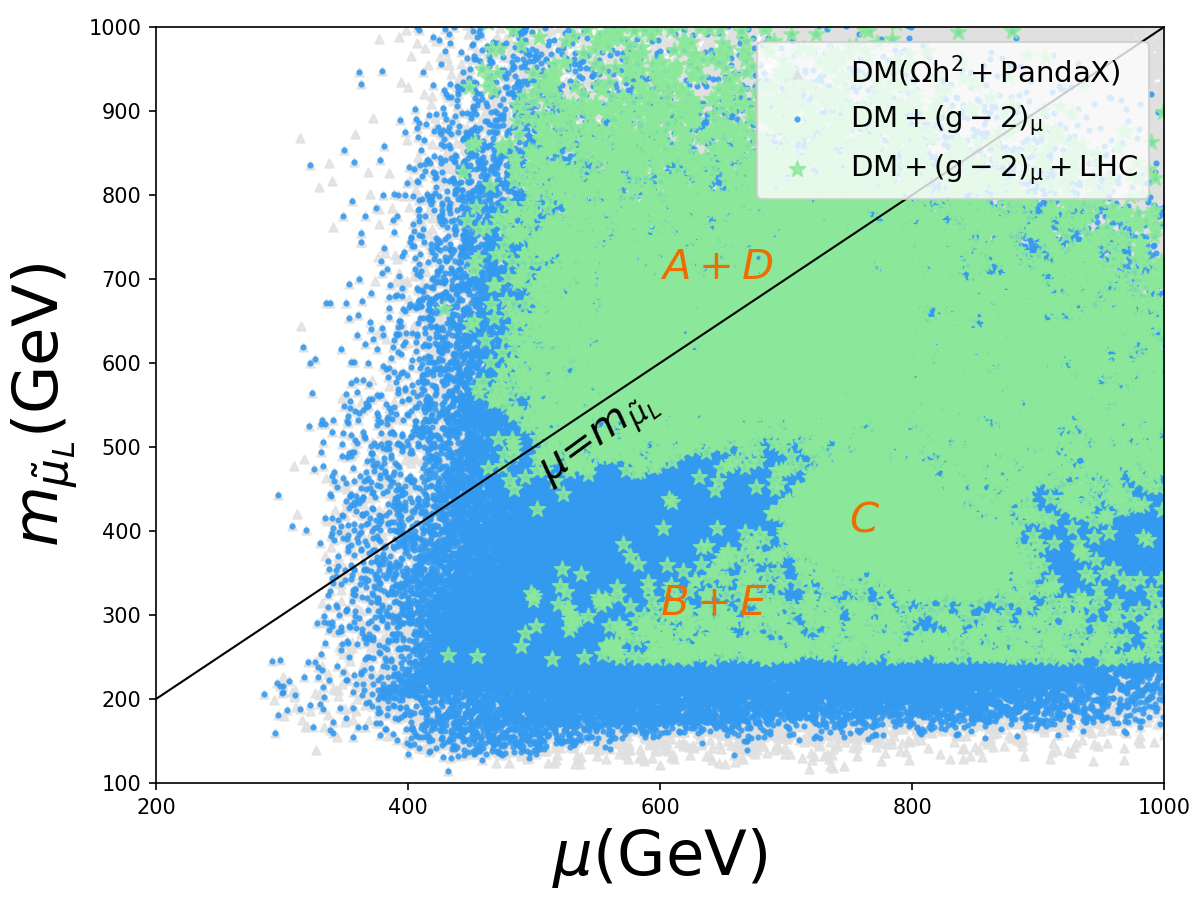}%\hspace{-0.3cm}
	\includegraphics[width=0.45\textwidth]{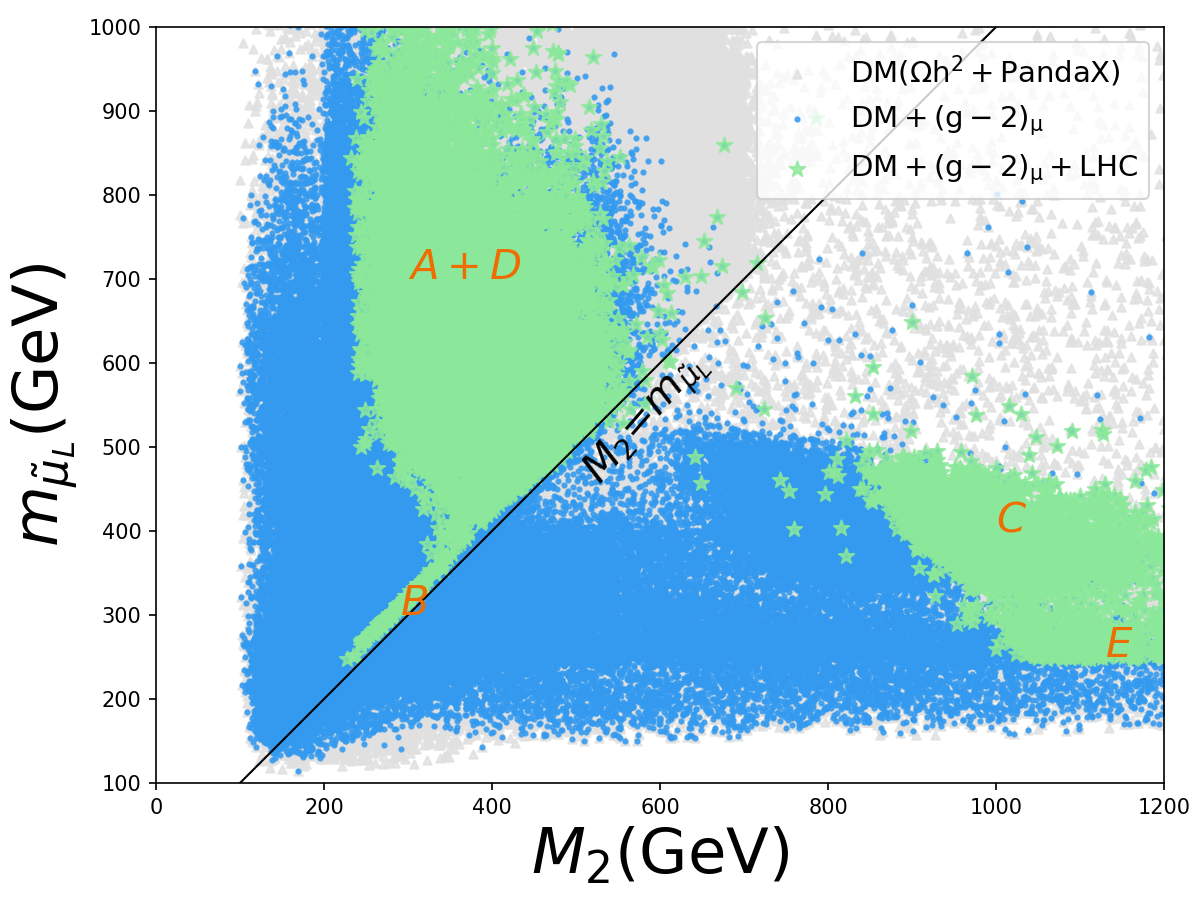}

	\caption{\label{fig2}
		Similar to Fig.~\ref{fig1}, but showing the correlations of the parameters that $a_\mu^{\rm SUSY}$ is sensitive to. $m_{\tilde{\mu}_L}$ is the mass of the left-handed-dominated smuon, which may significantly differ from the soft-breaking parameter $M_{\tilde{\mu}_L}$  defined at $Q=1~{\rm TeV}$ in Table~\ref{Table3}. Samples surviving the LHC restrictions are classified into five types, marked by A, B, C, D, and E in this figure. They are distinguished by different DM annihilation mechanisms and locations in the SUSY parameter space (see Table~\ref{Table4}).  }
\end{figure}

\begin{table}[tpb]
\centering
\caption{\label{Table4} Parameter spaces for the five types of samples in Fig.~\ref{fig2}, where the theoretical inputs in the last five columns are in units of GeV. The second column denotes which particles $\tilde{\chi}_1^0$ will co-annihilate with to acquire the measured DM density. In this aspect, Type-A and Type-B samples are different in that
$m_{\tilde{\mu}_L} \gtrsim M_2 + 30~{\rm GeV}$ for the former and $m_{\tilde{\mu}_L} \simeq M_2$ for the latter.  }

\vspace{0.2cm}

\resizebox{0.95 \textwidth}{!}{
\begin{tabular}{|c|c|c|c|c|c|c|}
\hline
Sample type & Annihilation partner & $|M_1|$ &$M_2$ & $\mu$ & $m_{\tilde{\mu}_L}$ & $m_{\tilde{\mu}_R}$ \\ \hline
Type-A & $\tilde{W}$ & (220, 560) &(230, 600) & (430, 1000) & (350, 1000) & (300, 1000) \\ \hline
Type-B & $\tilde{\mu}_L$ and $\tilde{W}$ & (210, 550) & (230, 600) & (430, 1000) & (240, 600) & (300, 1000) \\  \hline
Type-C & $\tilde{\mu}_L$ & (230, 540) & (600, 1400) & (540, 1000) & (240, 550) & (230, 1000) \\   \hline
Type-D & $\tilde{\mu}_R$ &(210, 520) &(300, 700) & (440, 1000) &  (400, 1000) & (210, 1000) \\  \hline
Type-E & $\tilde{\mu}_R$ &(230, 320) & (950, 1300) & (570, 1000) & (240, 600) & (230, 1000) \\ \hline
\end{tabular}}
\end{table}

Second, we concentrate on the interplay between the muon g-2 anomaly and the LHC restrictions. As introduced in the last section, explaining the muon g-2 anomaly requires more than one sparticle to be moderately light~\cite{Chakraborti:2020vjp}. In particular, $M_2$, $\mu$, and $m_{\tilde{\mu}_L}$ can not be very large simultaneously since the WHL contribution is usually dominant. This situation leads to sizable SUSY signals and thus strengthens the LHC restrictions. In Fig.~\ref{fig2}, we show the correlations of any two of the three parameters, $M_2$, $\mu$, and $m_{\tilde{\mu}_L}$, and also the correlation between $\mu$ and $\tan \beta$. The following distinct features are shown:
\begin{itemize}
\item The LHC restrictions have set lower bounds on the SUSY parameters: $\tan \beta \gtrsim 12$, $\mu \gtrsim 400~{\rm GeV}$, $M_2 \gtrsim 230~{\rm GeV}$, $m_{\tilde{\mu}_L} \gtrsim 240~{\rm GeV}$, and as shown in Fig.~\ref{fig1}, $|M_1| \gtrsim 210~{\rm GeV}$.
\item As indicated by the top left panel, the LHC restrictions are particularly strong for $\tan \beta \lesssim 25$. The underlying reason is the muon g-2 anomaly prefers light winos, higgsinos, and left-handed-dominant smuon as $\tan \beta$ becomes small.

\item In the case that $\tilde{\mu}_L$ is lighter than winos and/or higgsinos, the LHC restrictions are also strong, which is reflected by the wedge-shaped excluded regions on the $M_2-m_{\tilde{\mu}_L}$ and $\mu-m_{\tilde{\mu}_L}$ planes. This was because the heavy electroweakinos could decay into the slepton first and thus enhance the leptonic signal of the electroweakino pair production processes (compared with the case where $\tilde{\mu}_L$ is heavier than the electroweakinos). We elaborate on this point by fixing $m_{\tilde{\mu}_L} = 300~{\rm GeV}$ and varying $M_2$. For $M_2 \simeq 300~{\rm GeV}$, although the total wino pair production cross sections exceeded $550~{\rm fb}$~\cite{Fuks:2012qx,Fuks:2013vua}, there were some samples surviving the LHC constraints due to the small mass splittings between the wino-like particles and $\tilde{\chi}_1^0$. In this case, the wino-like particles decayed into $\tilde{\chi}_1^0$ and a soft virtual $Z$ or $W$, which made the signal detection difficult. With the increase in $M_2$, all the samples were excluded because the wino-like particles were copiously produced at the LHC due to their moderate lightness, and simultaneously the branching ratios of their decays into the slepton were sizable. Specifically, we found that the ratios were always larger than $20\%$. With the further increase in $M_2$, the wino pair production rates rapidly decreased so that the LHC constraints were weakened, reflected by the appearance of the green areas at $M_2 \simeq 900~{\rm GeV}$ in the bottom right panel. We add that this discussion can be applied to the $\mu-m_{\tilde{\mu}_L}$ plane in the bottom left panel.

\item Samples consistent with the LHC restrictions can be classified into five types, distinguished by their DM annihilation mechanisms and locations in the parameter space. They are marked as A, B, C, D, and E in the figure. In Table~\ref{Table4}, we provide the criteria of this classification. We will present benchmark points to reveal their properties and study the LHC restrictions later.

\end{itemize}

\begin{figure}[t]
	\centering
	\includegraphics[width=0.45\textwidth]{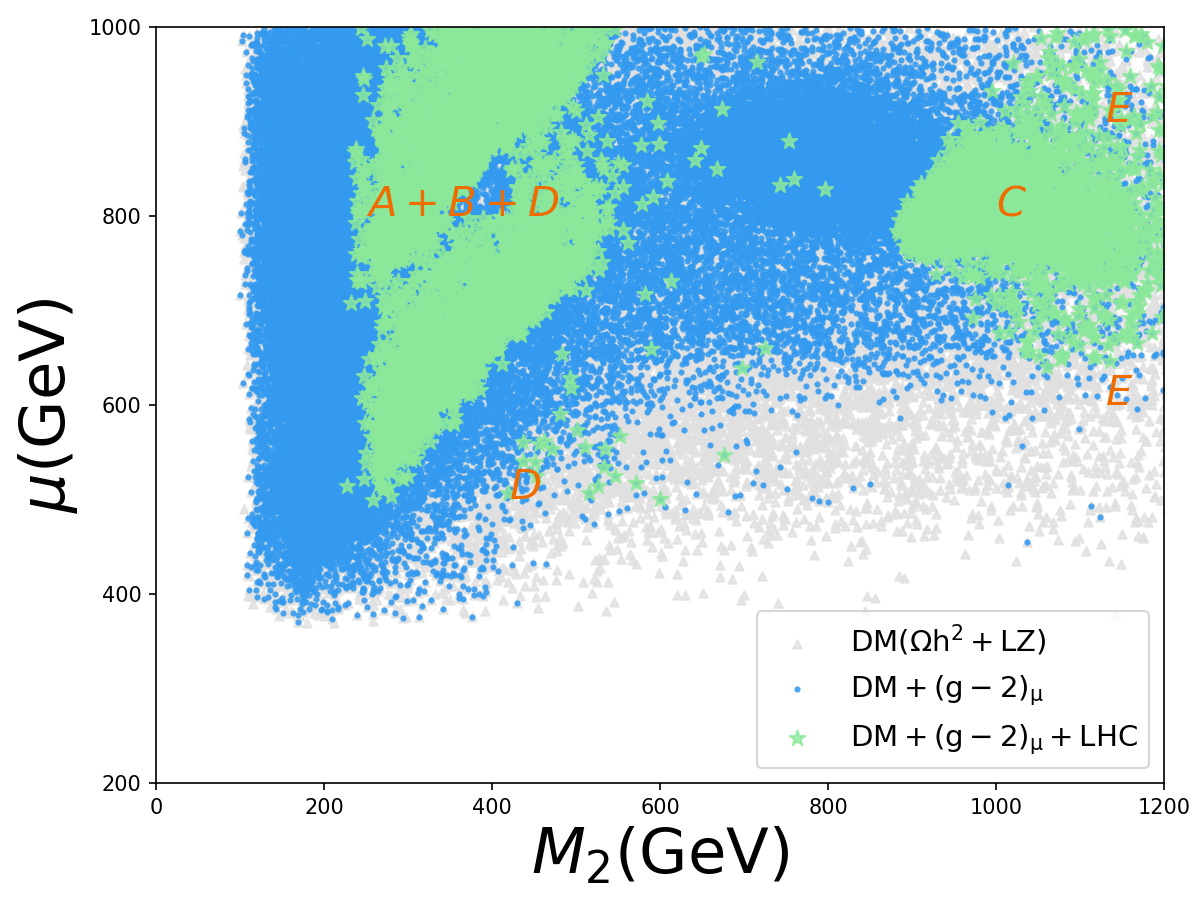}%\hspace{-0.3cm}
	\includegraphics[width=0.45\textwidth]{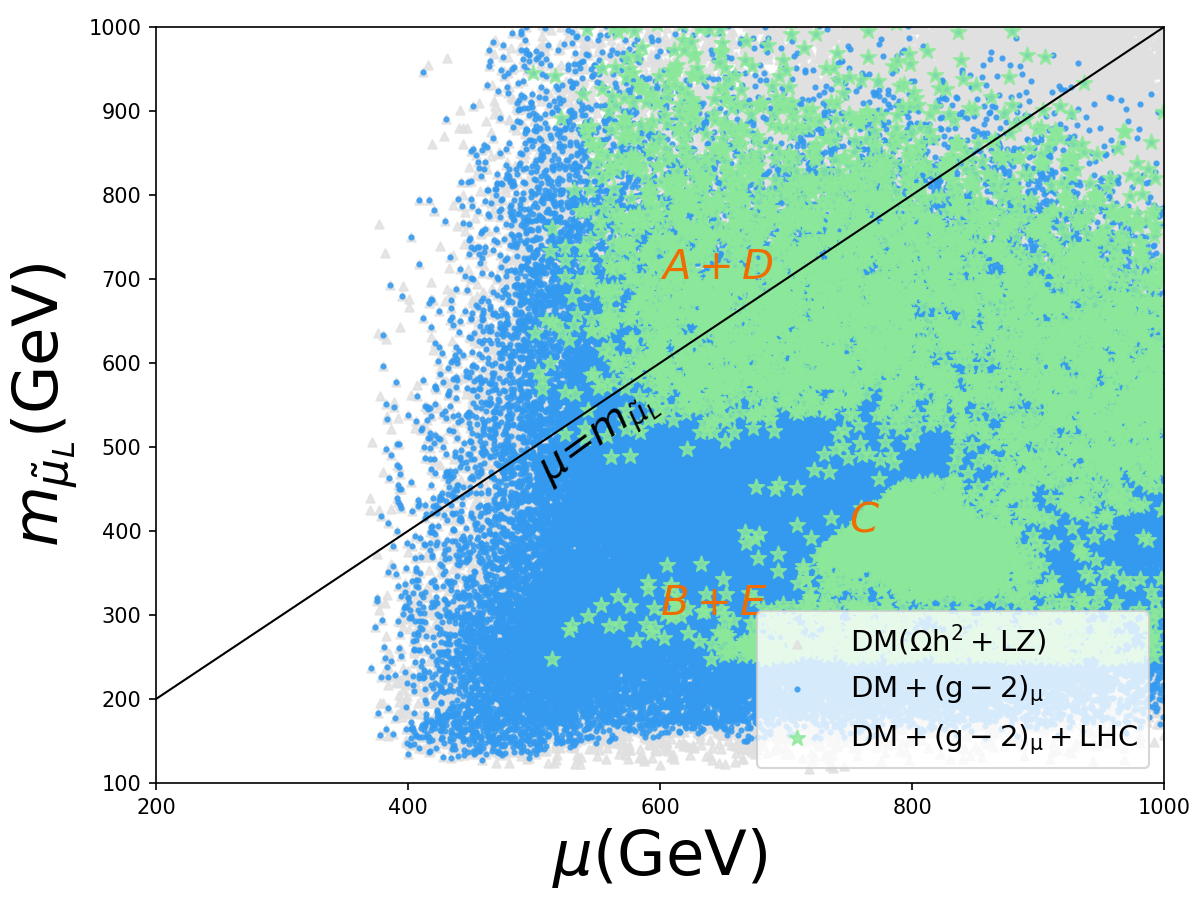}
	\caption{\label{fig3} Left panel: same as the upper right panel of Fig.~\ref{fig2}, except that all the samples were required further to satisfy the LZ restrictions. Right panel: same as the left panel of this figure, except that the samples were projected onto the $\mu-m_{\tilde{\mu}_L}$ plane.  }
\end{figure}

\begin{table}[tpb]
\centering
\caption{\label{Table5} Numbers of the samples studied by the simulations. They were categorized by the dominant annihilation mechanisms of the DM, which limit of the DM direct detection experiments was set, and whether the LHC restrictions were included in the research.  }

\vspace{0.3cm}

\begin{tabular}{c|cc|cc}
\hline
Annihilation Mechanisms & \multicolumn{2}{c|}{Before LHC Constraints}   & \multicolumn{2}{c}{After LHC Constraints}     \\ \hline
   & \multicolumn{1}{c|}{PandaX} &  LZ & \multicolumn{1}{c|}{PandaX} & LZ \\ \hline
 All  & \multicolumn{1}{c|}{58242} & 39657 & \multicolumn{1}{c|}{11204} & 5656  \\
 {$\tilde{B}-\tilde{W}$ co-annihilation}  & \multicolumn{1}{c|}{31108} & 20123 & \multicolumn{1}{c|}{7927} & 3435 \\
{$\tilde{B}-\tilde{\mu}_L -\tilde{W}$ co-annihilation}  & \multicolumn{1}{c|}{10052 } & 6622  & \multicolumn{1}{c|}{287 } & 174 \\
 {$\tilde{B}-\tilde{\mu}_L$ co-annihilation} & \multicolumn{1}{c|}{13445} & 11052  & \multicolumn{1}{c|}{2737} & 1884 \\
 {$\tilde{B}-\tilde{\mu}_R$ co-annihilation} & \multicolumn{1}{c|}{2869} & 1860  & \multicolumn{1}{c|}{253} & 163  \\
 {$Z-$ funnel} & \multicolumn{1}{c|}{408} & 0 & \multicolumn{1}{c|}{0} & 0 \\
 {$h-$ funnel}  & \multicolumn{1}{c|}{360} & 0 & \multicolumn{1}{c|}{0} & 0 \\ \hline
\end{tabular}
\end{table}

Third, we studied the impact of the LZ experiment on the MSSM. In Figs.~\ref{fig1} and \ref{fig2}, the PandaX-4T results were used to set upper bounds on the SI cross sections of the DM-nucleon scattering~\cite{PandaX-4T:2021bab}. We utilized the LZ limits to refine the samples further, projected the selected samples onto various panels, and compared the resulting figures with their correspondence plotted with the samples in Fig.~\ref{fig1}.
The most remarkable change came from the fact that $\mu$ was more strongly limited, which was reflected in the following aspects.
\begin{itemize}
\item Given the measured DM density, the LZ experiment alone required $\mu \gtrsim 380~{\rm GeV}$ for $M_1 < -100~{\rm GeV} $ and $\mu \gtrsim 600~{\rm GeV}$ for $M_1 > 100~{\rm GeV}$. If the restrictions from the muon g-2 anomaly and the LHC experiment were also included, the lower bounds became about $500~{\rm GeV}$ and $630~{\rm GeV}$, respectively, indicating that the theory needs a tuning of ${\cal{O}}(1\%)$ to predict the $Z$-boson mass~\cite{Baer:2012uy}. Compared with the restrictions from the PandaX-4T experiment, these bounds were improved by about $100~{\rm GeV}$.
\item The $Z$-mediated resonant annihilation became less favored because an enhanced $\mu$ reduced $C_{\tilde{\chi}_1^0 \tilde{\chi}_1^0 Z}$ in Eq.~\ref{SD-2}, and $2 |m_{\tilde{\chi}_1^0}|$ should be closer to $m_Z$ to obtain the measured density. This situation required the fine-tuning quantity defined in Eq. (19) of Ref.~\cite{Cao:2018rix} to be larger than about 150 to achieve the measured density. This conclusion also applied to the $h$-mediated resonant annihilation. We point out that, due to the tuning, these resonant annihilation scenarios were usually missed in the scans (see, e.g., the results in Fig.~\ref{fig4} of this study). This point was discussed with Bayesian statistics in footnote 6 of Ref.~\cite{Cao:2022ovk}.
\item Since the LZ constraint and the LHC restriction were sensitive to different SUSY parameters, they complemented each other in exploring the features of the MSSM. This was particularly so if one intended to explain the muon g-2 anomaly at the $2 \sigma$ level. The basic reason was that $\mu$ correlated with the other parameters by the anomaly, and any enhancement of $\mu$ in a massive higgsino scenario would make winos and $\tilde{\mu}_L$ lighter to keep $a_\mu^{\rm SUSY}$ unchanged. This situation usually improved the LHC restrictions.

    To show the combined effects, we projected the samples passing the LZ restrictions onto the $M_2-\mu$ and $\mu-m_{\tilde{\mu}_L}$ planes in Fig.~\ref{fig3} and compared the resulting panels with their corresponding ones in Fig.~\ref{fig2}. We also focused on the samples studied by the Monte Carlo simulations. We classified them by the dominant annihilation mechanisms of the DM, which limit of the DM experiments was set, and whether the LHC restrictions were included in this research. We presented the results in Table~\ref{Table5}. Both the figure and the table showed that the two experiments promoted each other to limit the parameter space of the MSSM.
\end{itemize}

\begin{figure}[t]
	\centering
	\includegraphics[width=0.45\textwidth]{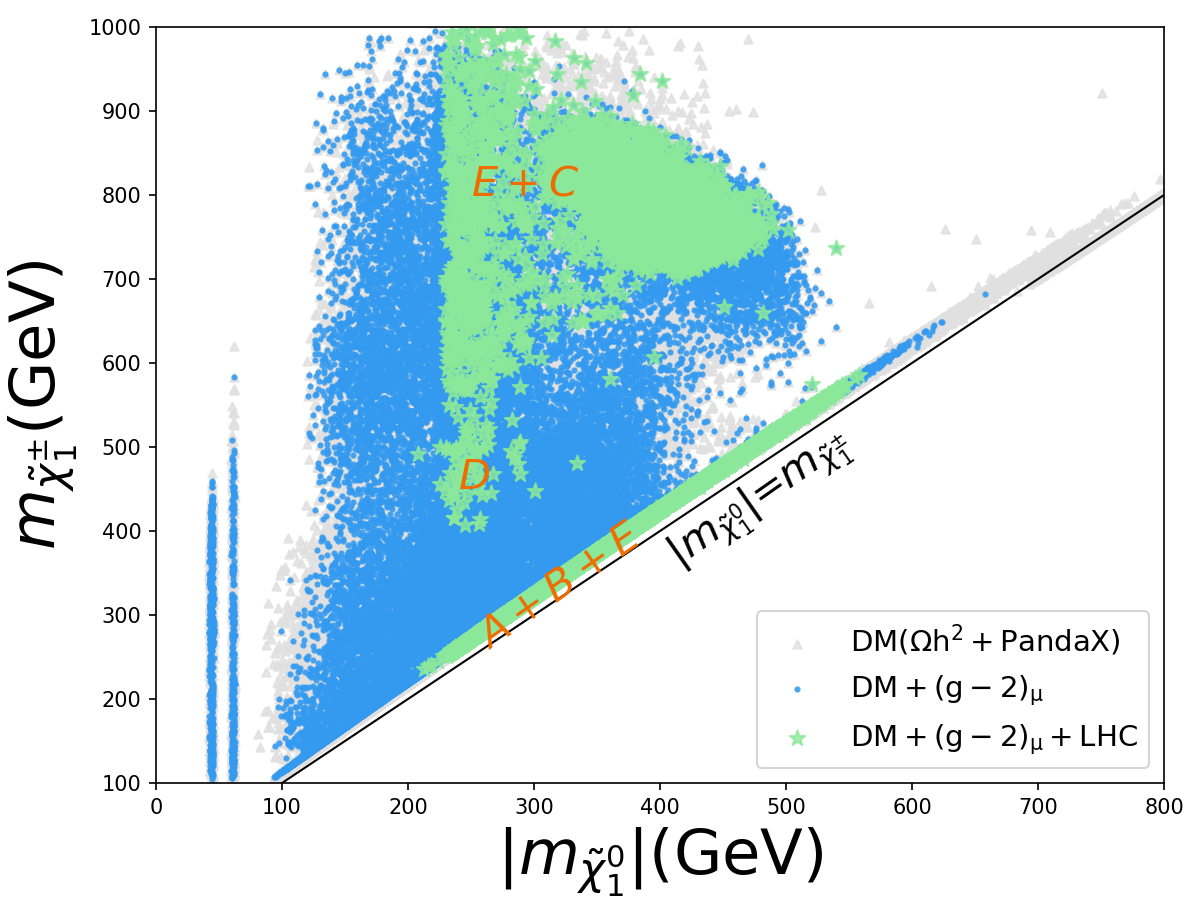}%\hspace{-0.3cm}
	\includegraphics[width=0.45\textwidth]{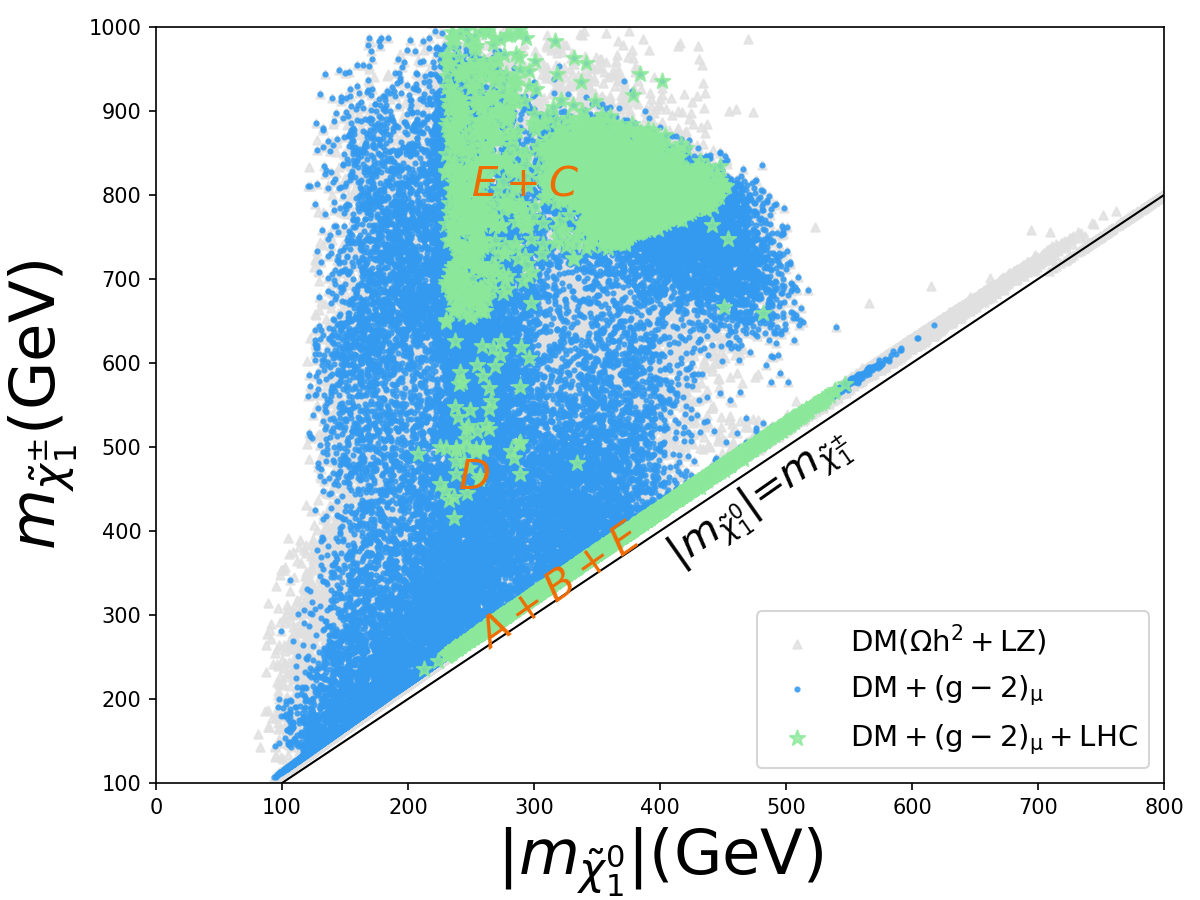}%\hspace{-0.3cm}

	\caption{\label{fig4} Distribution of $m_{\tilde{\chi}_1^\pm}$ versus $|m_{\tilde{\chi}_1^0}|$. The left panel studies the samples in Figs.~\ref{fig1}, while the right panel focuses on those in Fig.~\ref{fig3}. The classification in Table~\ref{Table4} was applied to the samples of this figure to illuminate the underlying physics. }
\end{figure}

\begin{figure}[t]
	\centering
	\includegraphics[width=0.45\textwidth]{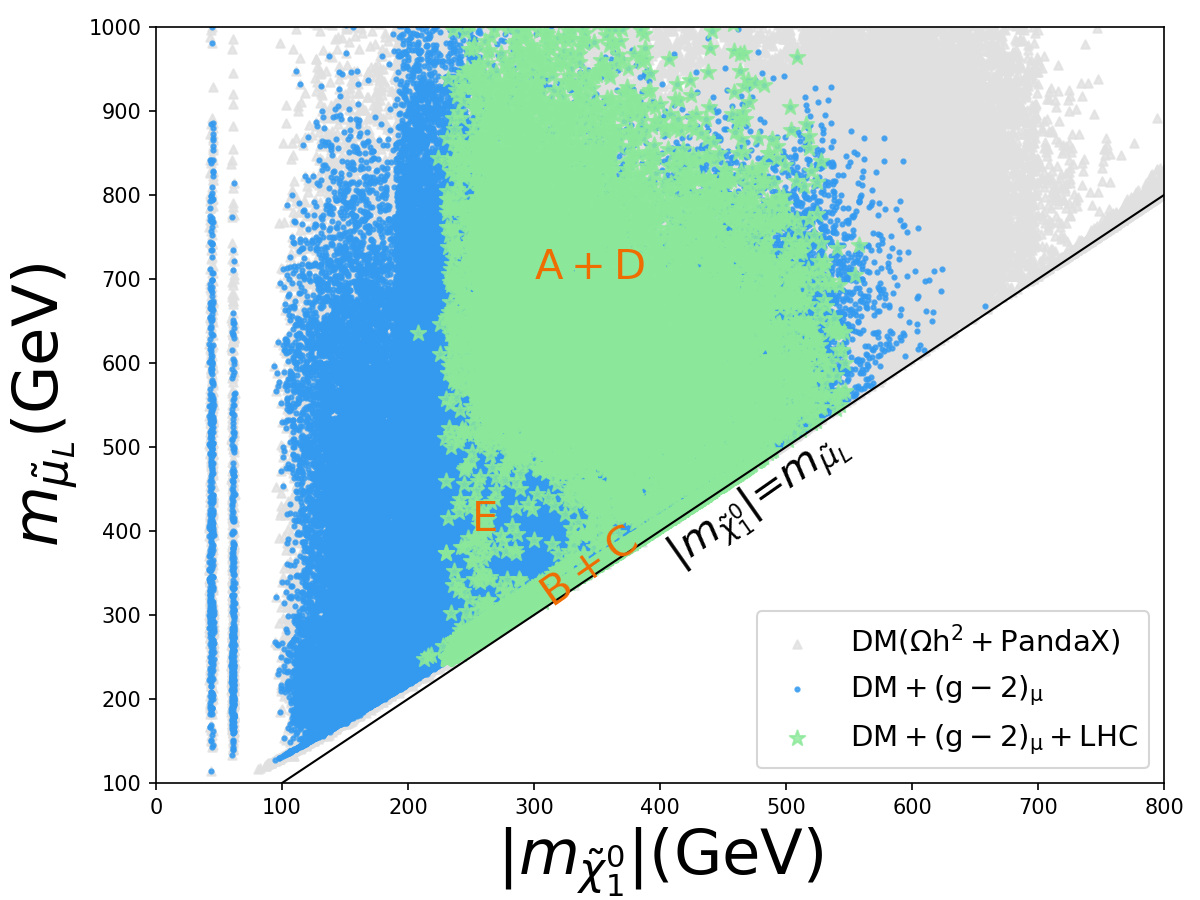}%\hspace{-0.3cm}
	\includegraphics[width=0.45\textwidth]{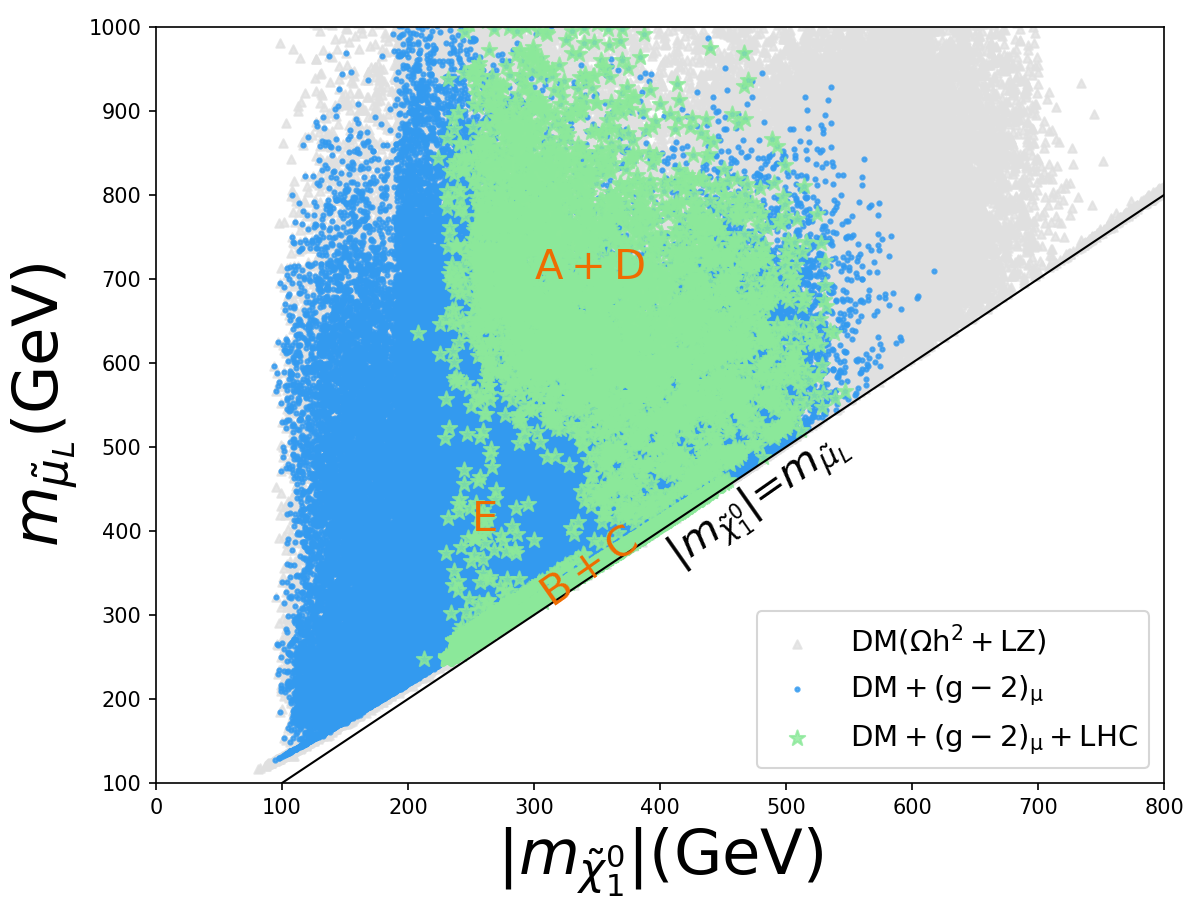}%\hspace{-0.3cm}
	\caption{\label{fig5} Same as Fig.~\ref{fig4} except that it shows the distribution on the $|m_{\tilde{\chi}_1^0}|-m_{\tilde{\mu}_L}$ plane.  }
\end{figure}

\begin{figure}[t]
	\centering
	\includegraphics[width=0.45\textwidth]{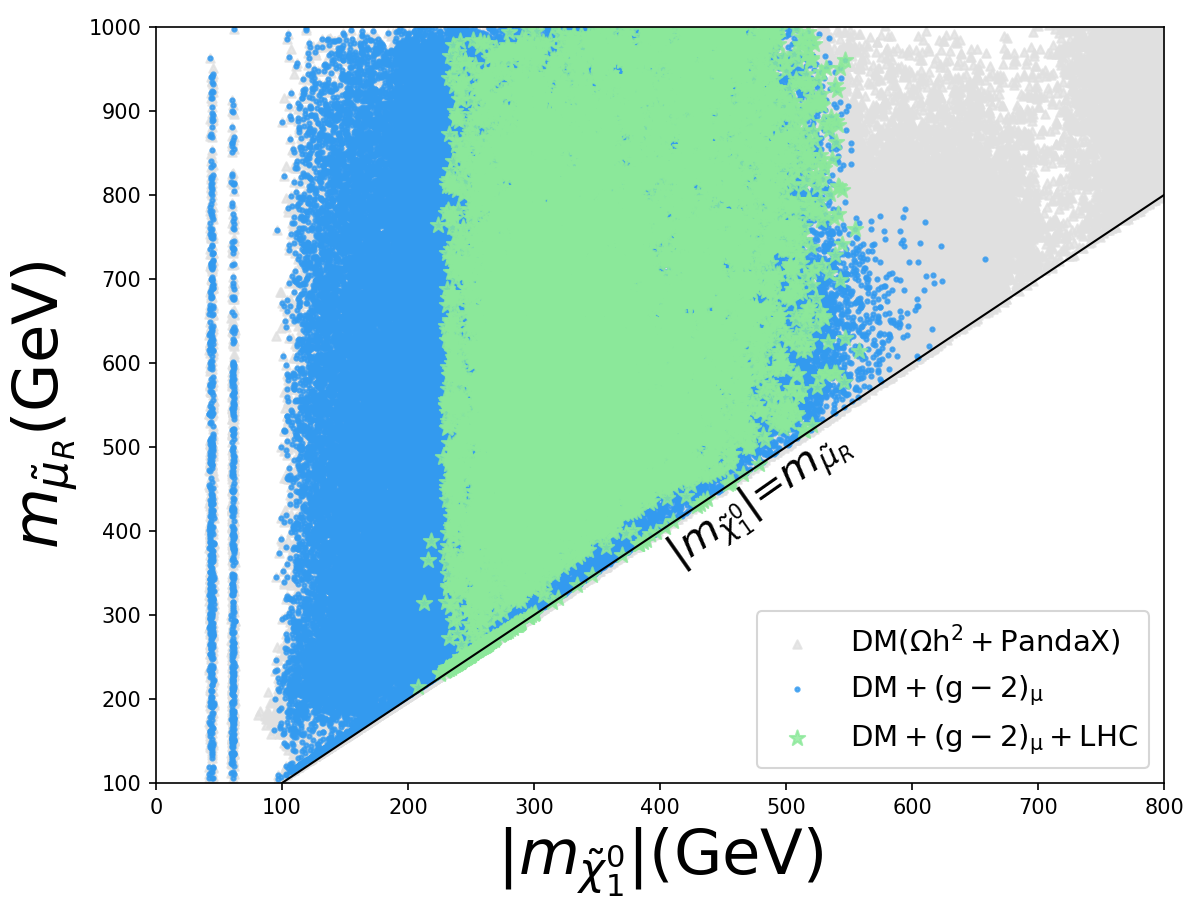}%\hspace{-0.3cm}
	\includegraphics[width=0.45\textwidth]{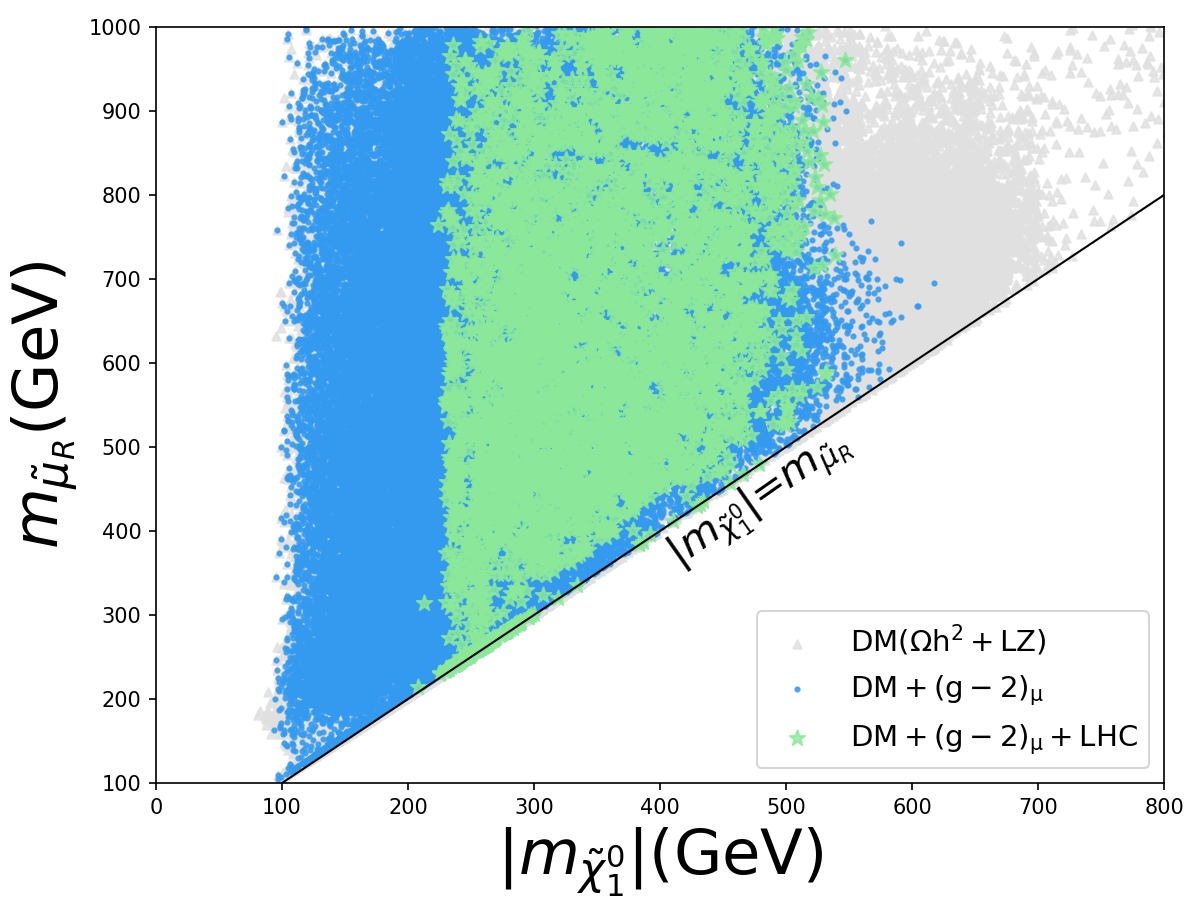}%\hspace{-0.3cm}
	\caption{\label{fig6} Same as Fig.~\ref{fig4} except that the samples are displayed on the $|m_{\tilde{\chi}_1^0}|-m_{\tilde{\mu}_R}$ plane.  }
\end{figure}

\begin{figure}[t]
	\centering
	\includegraphics[width=0.8\textwidth]{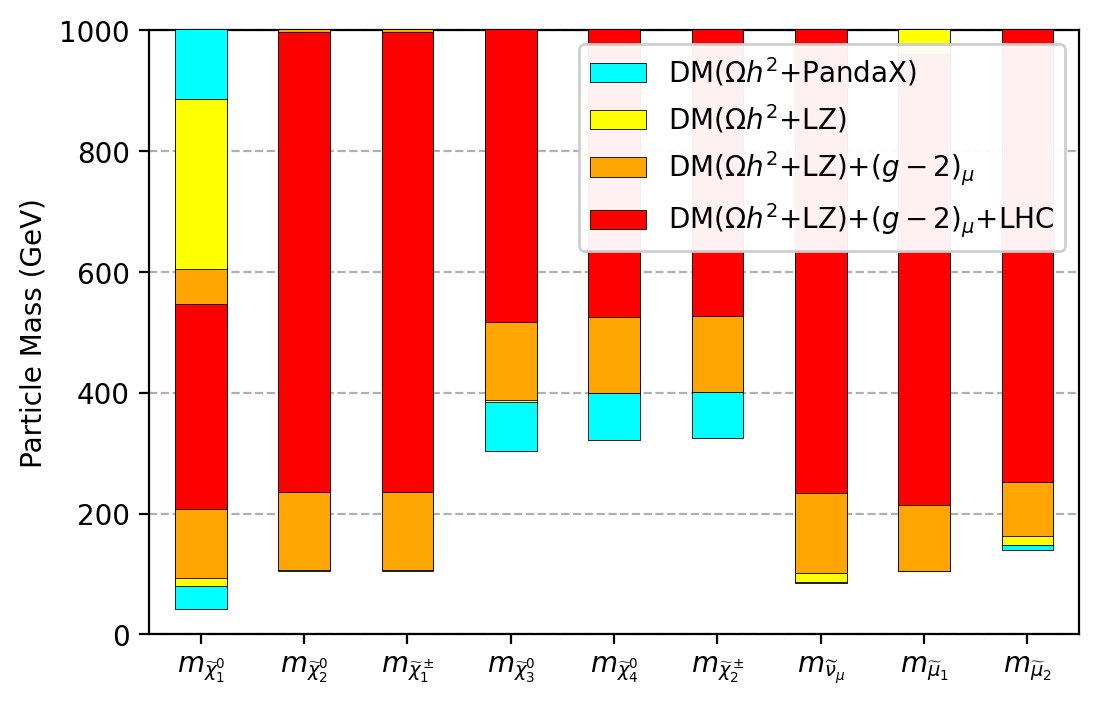}\hspace{-0.3cm}
	\caption{\label{fig7} Sparticle mass spectra preferred by different experiments. The cyan band was acquired from the gray samples in Fig.~\ref{fig1}, and the yellow, orange, and red bands were from the gray, blue, and green samples in Fig.~\ref{fig3}, respectively.  }
\end{figure}

Fourth, we surveyed the influences of these experiments on sparticle properties. We projected the samples of Fig.~\ref{fig1} (Fig.~\ref{fig3}) onto the $|m_{\tilde{\chi}_1^0}|-m_{\tilde{\chi}_1^\pm}$, $|m_{\tilde{\chi}_1^0}|-m_{\tilde{\mu}_L}$, and $|m_{\tilde{\chi}_1^0}|-m_{\tilde{\mu}_R}$ planes to acquire the left (right) panels of Figs.~\ref{fig4}, \ref{fig5}, and \ref{fig6}, respectively. The classification of the samples was the same as before. From these plots, we obtained the following points:
\begin{itemize}
	
  \item With the increase in $|m_{\tilde{\chi}_1^0}|$, the upper bounds of $m_{\tilde{\chi}_1^\pm}$ and $m_{\tilde{\mu}_L}$ decreased, and they terminated at $m_{\tilde{\chi}_1^\pm} \simeq 600~{\rm GeV}$ and $m_{\tilde{\mu}_L} \simeq 700~{\rm GeV}$ for  $m_{\tilde{\chi}_1^0} \simeq 570~{\rm GeV}$. This tendency was not evident for $m_{\tilde{\mu}_R}$ because $a_\mu^{\rm SUSY}$ was more sensitive to $M_2$, $\mu$, and $m_{\tilde{\mu}_L}$ than to $m_{\tilde{\mu}_R}$, as indicated by Eqs. \ref{eq:WHL}--\ref{eq:BLR}. 
  \item The LHC restrictions have set lower bounds on the sparticle mass spectra, which were $m_{\tilde{\chi}_1^0} \gtrsim 210~{\rm GeV}$, $m_{\tilde{\chi}_1^\pm} \gtrsim 235~{\rm GeV}$, $m_{\tilde{\mu}_L} \gtrsim 240~{\rm GeV}$, and $m_{\tilde{\mu}_R} \gtrsim 215~{\rm GeV}$ in this research. The basic reason for this phenomenon is as follows: if $\tilde{\chi}_1^0$ is lighter, more missing momentum will be emitted in the sparticle production processes at the LHC, which can improve the sensitivities of the experimental analyses; on the other hand, if sparticles other than $\tilde{\chi}_1^0$ are lighter, they will be more copiously produced at the LHC to increase the number of events containing multiple leptons. We emphasize that these bounds should be regarded as rough estimates, instead of accurate values, since far from enough samples were studied,
      given the broad parameter space of the MSSM.
\end{itemize}

Finally, we summarize the sparticle mass spectra preferred by different experiments in Fig.~\ref{fig7}. This figure reveals that $ 570~{\rm GeV} \gtrsim m_{\tilde{\chi}_1^0} \gtrsim 210~{\rm GeV}$, $m_{\tilde{\chi}_2^0}, m_{\tilde{\chi}_1^\pm} \gtrsim 235~{\rm GeV}$, $m_{\tilde{\chi}_3^0} \gtrsim 515~{\rm GeV}$, $m_{\tilde{\chi}_4^0} \gtrsim 525~{\rm GeV}$, $m_{\tilde{\chi}_2^\pm} \gtrsim 530~{\rm GeV}$, $m_{\tilde{\nu}_\mu} \gtrsim 235~{\rm GeV}$, $ 950~{\rm GeV} \gtrsim m_{\tilde{\mu}_1} \gtrsim 215~{\rm GeV}$, and $m_{\tilde{\mu}_2} \gtrsim 250~{\rm GeV}$ if all the latest restrictions are considered. The lower bounds come from the LHC restrictions, and the upper bounds arise from the explanation of the muon g-2 anomaly at the $2 \sigma$ level. In addition, it was verified that $\tilde{\chi}_{2}^0$ and $\tilde{\chi}_1^\pm$ were wino-dominated when they were lighter than about $500~{\rm GeV}$ and $m_{\tilde{\mu}_1}$ might be either $m_{\tilde{\mu}_L}$ or $m_{\tilde{\mu}_R}$.  It was also verified that $ 600~{\rm GeV} \gtrsim m_{\rm NLSP} \gtrsim 220~{\rm GeV}$ and $ 700~{\rm GeV} \gtrsim m_{\rm NNLSP} \gtrsim 250~{\rm GeV}$, where the NLSP and NNLSP might be either electroweakinos or sleptons.
In principle, the sparticles other than $\tilde{\chi}_{1}^0$ and $\tilde{\mu}_{1}^0$ are also upper bounded in mass by the explanation of the anomaly. We do not provide these bounds since the parameter space in Table~\ref{Table3} is limited.

\begin{table}[tpb]
\centering
\caption{\label{Table6} Supplement to Table~\ref{Table5} with the SRs that contribute to the largest $R$-value and their capability to exclude the samples, expressed by the percentage of the total numbers in the second column. This table only counts the samples satisfying the LZ restrictions and studied by the simulations. The second and third columns denote the sample numbers before and after implementing the LHC restrictions, respectively, which are also presented in Table~\ref{Table5}. They reflect that the LHC restrictions on the $\tilde{B}-\tilde{\mu}_L-\tilde{W}$ and $\tilde{B}-\tilde{\mu}_R$ co-annihilation cases are very strong. One can understand this feature from the previous discussions in this study and the benchmark points listed below.
The fourth column reveals that only \texttt{SR-1} from the experimental analyses in Ref.~\cite{CMS:2020bfa} and \texttt{SR-2} from Ref.~\cite{CMS:2017moi,CMS:2018szt} played a role in excluding the $\tilde{B}-\tilde{\mu}_L$ co-annihilation case, which was different from the other cases.
}

\vspace{0.2cm}
\resizebox{\textwidth}{!}{
\begin{tabular}{c|c|c|c}
\hline
Annihilation mechanism    & Before & After & SRs and their exclusion percentages                    \\ \hline
All & 39657         & 5656         & SR-1(23.7\%),SR-2(18.5\%),SR-3(12.7\%),SR-4(11.6\%) \\
$\tilde{B} - \tilde{W}$ co-annihilation        & 20123         & 3435         & SR-3(23.8\%), SR-4(21.1\%), SR-5(6.3\%), SR-1(0.1\%)  \\
$\tilde{B} - \tilde{\mu}_L - \tilde{W}$ co-annihilation       & 6622         & 174         & SR-3(28.6\%),SR-4(16.4\%),SR-2(14.4\%),SR-5(11.7\%)  \\
$\tilde{B} - \tilde{\mu}_L$ co-annihilation       & 11052         & 1884         & SR-1(52.0\%),SR-2(30.4\%)  \\
$\tilde{B} - \tilde{\mu}_R$ co-annihilation       & 1860          & 163          & SR-2(44.0\%), SR-4(32.3\%), SR-3(6.2\%), SR-1(0.8\%)   \\ \hline
\end{tabular}}
\end{table}

\subsection{More details of LHC restrictions}

Even in the simple realization of SUSY, such as the MSSM, the decay products of heavy sparticles are complex (see the benchmark points presented below). As a result, the pair productions of winos, higgsinos, $\tilde{\mu}_L$, and $\tilde{\mu}_R$ at the LHC, $p p \to \bar{\tilde{W}} \tilde{W}, \bar{\tilde{H}} \tilde{H}, \tilde{\mu}_L^\ast \tilde{\mu}_L, \tilde{\mu}_R^\ast \tilde{\mu}_R$, may contribute to the same SR of the analyses in Tables~\ref{Table1} and \ref{Table2}. All these contributions must be suppressed for any parameter point to circumvent the LHC restrictions, which may occur in the following situations:
\begin{enumerate}
\item The DM candidate is massive, compared with the results of pertinent experimental analyses. In this case, all SM particles in the final state are not energetic enough, and the missing momentum emitted in the sparticle production processes tends to be small.
\item The mass splitting between the decaying sparticle and $\tilde{\chi}_1^0$ is less than several tens of GeV. In this compressed spectra case, the SM particles as the decay product are soft and hard to detect without a deliberate search strategy.
\item Heavy sparticles decay by several channels with comparable branching ratios in terms of size. This situation usually leads to complicated final decay states.
\item Sparticles are sufficiently heavy, compared with their experimental exclusion bounds in simplified models, so that their production cross sections are negligibly small.
 \end{enumerate}
We refer to these situations as survival mechanisms I, II, III, and IV in the following discussion. An illuminating example of situations 1 and 2 was presented in Fig. 16 of Ref.~\cite{ATLAS:2021moa}, which concluded that there were no LHC restrictions on winos in the $\tilde{B}-\tilde{W}$ co-annihilation case if $m_{\tilde{\chi}_1^0} \gtrsim 220~{\rm GeV}$. Evidently, this bound is significantly weaker than the other searches for winos at the LHC.

\begin{table}[tpb]
	\caption{\label{Table7} Two benchmark points, P1 and P2, for Type-A and -B samples in Table \ref{Table4}, respectively. Both points satisfy all the restrictions listed in the text. The $R$-value of each point and its corresponding SR are presented in the last line of this table.}
	\vspace{0.3cm}
	\resizebox{1.0\textwidth}{!}{
		\begin{tabular}{llll|llll}
			\hline\hline
			\multicolumn{4}{l|}{\texttt{Type-A BP P1}}& \multicolumn{4}{l}{\texttt{Type-B BP P2}}                                                                        \\ \hline
			\multicolumn{1}{l}{$\mu$}     & \multicolumn{1}{r}{750.7 GeV} & \multicolumn{1}{l}{$\text{m}_\text{h}$}         &\multicolumn{1}{r|}{125.5 GeV}    & \multicolumn{1}{l}{$\mu$}     & \multicolumn{1}{r}{636.8 GeV} & \multicolumn{1}{l}{$\text{m}_\text{h}$}         &\multicolumn{1}{r}{125.2 GeV}     \\
			\multicolumn{1}{l}{$\text{tan}\beta$}     & \multicolumn{1}{r}{53.2} & \multicolumn{1}{l}{$\text{m}_{\text{A}}$}        &\multicolumn{1}{r|}{10359.8 GeV}     & \multicolumn{1}{l}{$\text{tan}\beta$}     & \multicolumn{1}{r}{35.5} & \multicolumn{1}{l}{$\text{m}_{\text{A}}$}        &\multicolumn{1}{r}{7359.7 GeV}     \\
			\multicolumn{1}{l}{$\text{A}_t$}      & \multicolumn{1}{r}{2496.8 GeV} & \multicolumn{1}{l}{$\text{m}_{\text{H}^{\pm}}$}         &\multicolumn{1}{r|}{10337.7 GeV}     & \multicolumn{1}{l}{$\text{A}_t$}      & \multicolumn{1}{r}{2409.8 GeV} & \multicolumn{1}{l}{$\text{m}_{\text{H}^{\pm}}$}         &\multicolumn{1}{r}{7363.2 GeV}     \\
			\multicolumn{1}{l}{$\text{M}_1$}        & \multicolumn{1}{r}{$-380.0$ GeV} &  \multicolumn{1}{l}{$\text{m}_{\tilde{\nu}_{\mu}}$}        &\multicolumn{1}{r|}{678.0 GeV}    & \multicolumn{1}{l}{$\text{M}_1$}        & \multicolumn{1}{r}{$-292.9$ GeV} & \multicolumn{1}{l}{$\text{m}_{\tilde{\nu}_{\mu}}$}        &\multicolumn{1}{r}{316.2 GeV}     \\
			\multicolumn{1}{l}{$\text{M}_2$} & \multicolumn{1}{r}{386.9 GeV} &  \multicolumn{1}{l}{$\text{m}_{\tilde{\chi}_1^0}$}        &\multicolumn{1}{r|}{$-380.7$ GeV}    & \multicolumn{1}{l}{$\text{M}_2$} & \multicolumn{1}{r}{304.9 GeV} &  \multicolumn{1}{l}{$\text{m}_{\tilde{\chi}_1^0}$}        &\multicolumn{1}{r}{$-292.7$ GeV}    \\		
		   \multicolumn{1}{l}{$\text{M}_{\tilde{\mu}_{\text{L}}}$}        & \multicolumn{1}{r}{550.2 GeV} & \multicolumn{1}{l}{$\text{m}_{\tilde{\chi}_2^0}$}        &\multicolumn{1}{r|}{404.8 GeV}     &  \multicolumn{1}{l}{$\text{M}_{\tilde{\mu}_{\text{L}}}$}        & \multicolumn{1}{r}{188.4 GeV} &   \multicolumn{1}{l}{$\text{m}_{\tilde{\chi}_2^0}$}        &\multicolumn{1}{r}{317.4 GeV}   \\
               \multicolumn{1}{l}{$\text{M}_{\tilde{\mu}_{\text{R}}}$}        & \multicolumn{1}{r}{972.5 GeV} & \multicolumn{1}{l}{$\text{m}_{\tilde{\chi}_3^0}$}        &\multicolumn{1}{r|}{$-772.6$ GeV}     & \multicolumn{1}{l}{$\text{M}_{\tilde{\mu}_{\text{R}}}$}        & \multicolumn{1}{r}{708.0 GeV} & \multicolumn{1}{l}{$\text{m}_{\tilde{\chi}_3^0}$}        &\multicolumn{1}{r}{$-656.4$ GeV} \\
			\multicolumn{1}{l}{$a_{\mu}^{\text{SUSY}}$}        & \multicolumn{1}{r}{$1.95\times 10^{-9}$} & \multicolumn{1}{l}{$\text{m}_{\tilde{\chi}_4^0}$}        &\multicolumn{1}{r|}{777.7 GeV}     & \multicolumn{1}{l}{$a_{\mu}^{\text{SUSY}}$}        & \multicolumn{1}{r}{$2.43\times 10^{-9}$} & \multicolumn{1}{l}{$\text{m}_{\tilde{\chi}_4^0}$}        &\multicolumn{1}{r}{662.0 GeV} \\
			\multicolumn{1}{l}{${\Omega h}^2$}        & \multicolumn{1}{r}{0.10} & \multicolumn{1}{l}{$\text{m}_{\tilde{\chi}_1^{\pm}}$}        &\multicolumn{1}{r|}{405.0 GeV}     & \multicolumn{1}{l}{${\Omega h}^2$}        & \multicolumn{1}{r}{0.12} & \multicolumn{1}{l}{$\text{m}_{\tilde{\chi}_1^{\pm}}$}        &\multicolumn{1}{r}{317.6 GeV}     \\
			\multicolumn{1}{l}{$\sigma_{p}^{\text{SI}}$}       & \multicolumn{1}{r}{$4.50 \times 10^{-47}\text{cm}^{2}$} & \multicolumn{1}{l}{$\text{m}_{\tilde{\chi}_2^{\pm}}$}        &\multicolumn{1}{r|}{779.6 GeV}     & \multicolumn{1}{l}{$\sigma_{p}^{\text{SI}}$}       & \multicolumn{1}{r}{$4.34 \times 10^{-47}\text{cm}^{2}$} &  \multicolumn{1}{l}{$\text{m}_{\tilde{\chi}_2^{\pm}}$}      &\multicolumn{1}{r}{664.2 GeV}  \\
			\multicolumn{1}{l}{$\sigma_{p}^{\text{SD}}$}       & \multicolumn{1}{r}{$5.38 \times 10^{-43}\text{cm}^{2}$} &  \multicolumn{1}{l}{$\text{m}_{\tilde{\mu}_{\text{L}}}$}        & \multicolumn{1}{r|}{682.2 GeV}    & \multicolumn{1}{l}{$\sigma_{p}^{\text{SD}}$}       & \multicolumn{1}{r}{$9.21\times 10^{-43}\text{cm}^{2}$} &   \multicolumn{1}{l}{$\text{m}_{\tilde{\mu}_{\text{L}}}$}        & \multicolumn{1}{r}{325.6 GeV}   \\
			& &   \multicolumn{1}{l}{$\text{m}_{\tilde{\mu}_{\text{R}}}$}        & \multicolumn{1}{r|}{806.5 GeV}    & & &   \multicolumn{1}{l}{$\text{m}_{\tilde{\mu}_{\text{R}}}$}        & \multicolumn{1}{r}{612.9 GeV}   \\ \hline
			\multicolumn{2}{l}{$\text{N}_{11},\text{N}_{12},\text{N}_{13},\text{N}_{14}$}           & \multicolumn{2}{r|}{$-0.997,-0.004,-0.075,-0.036$}                 & \multicolumn{2}{l}{$\text{N}_{11},\text{N}_{12},\text{N}_{13},\text{N}_{14}$}           & \multicolumn{2}{r}{$-0.996,-0.004,-0.083,-0.036$}                 \\
			\multicolumn{2}{l}{$\text{N}_{21},\text{N}_{22},\text{N}_{23},\text{N}_{24}$}           & \multicolumn{2}{r|}{$0.004,\,~~0.987,-0.140,\,~~0.075$}                 &               \multicolumn{2}{l}{$\text{N}_{21},\text{N}_{22},\text{N}_{23},\text{N}_{24}$}           & \multicolumn{2}{r}{$0.006,\,~~0.985,-0.157,\,~~0.079$}                 \\
			\multicolumn{2}{l}{$\text{N}_{31},\text{N}_{32},\text{N}_{33},\text{N}_{34}$}           & \multicolumn{2}{r|}{$~0.079,\,-0.046,-0.702,-0.706$}                 &               \multicolumn{2}{l}{$\text{N}_{31},\text{N}_{32},\text{N}_{33},\text{N}_{34}$}           & \multicolumn{2}{r}{$~0.084,-0.055,-0.700,-0.707$}                 \\
			\multicolumn{2}{l}{$\text{N}_{41},\text{N}_{42},\text{N}_{43},\text{N}_{44}$}           & \multicolumn{2}{r|}{$-0.027,\,~~0.152,\,~~0.694,-0.703$}                 &               \multicolumn{2}{l}{$\text{N}_{41},\text{N}_{42},\text{N}_{43},\text{N}_{44}$}           & \multicolumn{2}{r}{$-0.033,\,~~0.166,\,~~0.691,-0.702$}                                  \\ \hline
			\multicolumn{2}{l}{Annihilations}                     & \multicolumn{2}{l|}{Fractions[\%]}        & \multicolumn{2}{l}{Annihilations}                     & \multicolumn{2}{l}{Fractions[\%]}        \\
			\multicolumn{2}{l}{$\tilde{B}-\tilde{W}$ Co-annihilation}                  & \multicolumn{2}{l|}{89.9}                 & \multicolumn{2}{l}{$\tilde{B}-\tilde{W}/\tilde{B}-\tilde{\mu}_L$ Co-annihilation}                  & \multicolumn{2}{l}{73.0/14.5}                 \\
			\hline
			\multicolumn{2}{l}{Decays}                            & \multicolumn{2}{l|}{Branching ratios[\%]} & \multicolumn{2}{l}{Decays}                            & \multicolumn{2}{l}{Branching ratios[\%]} \\
			\multicolumn{2}{l}{$\tilde{\chi}_2^0 \to \tilde{\chi}_1^0Z^*/\tilde{\chi}_1^0h^*$}& \multicolumn{2}{l|}{100}
			&\multicolumn{2}{l}{$\tilde{\chi}_2^0 \to \tilde{\nu}_{\mu}\nu_{\mu}$}& \multicolumn{2}{l}{100}                 \\
			\multicolumn{2}{l}{$\tilde{\chi}_3^0 \to \tilde{\chi}_1^{\pm} \text{W}^{\mp}/ \tilde{\chi}_2^0Z/ \tilde{\chi}_1^0h/\tilde{\chi}_2^0h/\tilde{\chi}_1^0Z $} & \multicolumn{2}{l|}{61.0/27.0/8.3/2.3/1.1}                 & \multicolumn{2}{l}{$\tilde{\chi}_3^0 \to \tilde{\chi}_1^{\pm} \text{W}^{\mp} / \tilde{\chi}_2^0Z/ \tilde{\chi}_1^0h / \tilde{\chi}_2^0h / \tilde{\chi}_1^0 Z$}                           & \multicolumn{2}{l}{61.2/26.4/7.7/2.6/1.4}                 \\
			\multicolumn{2}{l}{$\tilde{\chi}_4^0 \to \tilde{\chi}_1^{\pm}W^{\mp}/\tilde{\chi}_2^0h/\tilde{\chi}_1^0Z/\tilde{\chi}_2^0Z/\tilde{\chi}_1^0h$}& \multicolumn{2}{l|}{61.3/26.2/8.3/2.8/1.0}                 & \multicolumn{2}{l}{$\tilde{\chi}_4^0 \to \tilde{\chi}_1^{\pm}W^{\mp}/\tilde{\chi}_2^0 h/\tilde{\chi}_1^0 Z / \tilde{\chi}_2^0 Z/\tilde{\nu}_{\mu}\nu_{\mu}/\tilde{\chi}_1^0 h/ \tilde{\mu}^{\pm}_{\text{L}} \mu^{\mp}$}                           & \multicolumn{2}{l}{60.9/24.7/7.6/3.3/1.4/1.2/0.9}\\
			\multicolumn{2}{l}{$\tilde{\chi}_1^{\pm} \to \tilde{\chi}_1^0(W^{\pm})^*  $}                           & \multicolumn{2}{l|}{100}                 & \multicolumn{2}{l}{$\tilde{\chi}_1^{\pm} \to \tilde{\nu}_{\mu}\mu^{\pm} $}                           & \multicolumn{2}{l}{99.9}                 \\
			\multicolumn{2}{l}{$\tilde{\chi}_2^{\pm} \to \tilde{\chi}_2^0W^{\pm}/ \tilde{\chi}_1^{\pm}Z/ \tilde{\chi}_1^{\pm}h/ \tilde{\chi}_1^0W^{\pm}$}                     & \multicolumn{2}{l|}{31.0/30.1/28.6/9.9}                 & \multicolumn{2}{l}{$\tilde{\chi}_2^{\pm} \to \tilde{\chi}_2^0W^{\pm}/\tilde{\chi}_1^{\pm}Z/\tilde{\chi}_1^{\pm}h / \tilde{\chi}_1^0W^{\pm} / \tilde{\mu}^{\pm}_{\text{L}} \nu_{\mu} / \tilde{\nu}_{\mu}\mu^{\pm} $}                     & \multicolumn{2}{l}{30.8/29.7/27.4/9.6/1.9/0.6}                 \\
			\multicolumn{2}{l}{$\tilde{\mu}_{\text{L}}^{\pm} \to\tilde{\chi}_1^{\pm}\nu_{\mu}/\tilde{\chi}_2^0\mu^{\pm}/\tilde{\chi}_1^0\mu^{\pm}$}                              & \multicolumn{2}{l|}{58.8/30.0/11.2}                 & \multicolumn{2}{l}{$\tilde{\mu}_{\text{L}}^{\pm} \to\tilde{\chi}_1^0\mu^{\pm}/ \tilde{\chi}_1^{\pm}\nu_{\mu}/ \tilde{\chi}_2^0\mu^{\pm} $}                              & \multicolumn{2}{l}{63.4/23.7/12.9}                 \\
			\multicolumn{2}{l}{$\tilde{\mu}_{\text{R}}^{\pm} \to \tilde{\chi}_1^0\mu^{\pm}/ \tilde{\nu}_{\mu}\text{W}^{\pm}$}                              & \multicolumn{2}{l|}{99.8/0.1}                 & \multicolumn{2}{l}{$\tilde{\mu}_{\text{R}}^{\pm} \to \tilde{\chi}_1^0\mu^{\pm}/ \tilde{\nu}_{\mu} W^{\pm}/\tilde{\mu}^{\pm}_{\text{L}} h/ \tilde{\mu}^{\pm}_{\text{L}} Z $} &\multicolumn{2}{l}{99.5/0.3/0.1/0.1}\\
			\multicolumn{2}{l}{$\tilde{\nu}_{\mu} \to \tilde{\chi}_1^{\pm}\mu^{\mp}/\tilde{\chi}_2^0\nu_{\mu}/\tilde{\chi}_1^0\nu_{\mu}$}& \multicolumn{2}{l|}{59.8/29.4/10.8}                 & \multicolumn{2}{l}{$\tilde{\nu}_{\mu} \to \tilde{\chi}_1^{0}\nu_{\mu}$}                             & \multicolumn{2}{l}{100}                 \\	\hline
			\multicolumn{2}{l}{$R$ value}                           & \multicolumn{2}{l|}{0.37, \textbf{S−high−mm−05} in \texttt{SR-3}}                 & \multicolumn{2}{l}{$R$ value}                           & \multicolumn{2}{l}{0.30, \textbf{SRG07\_0j\_mll} in \texttt{SR-1}}                 \\
			\hline\hline
\end{tabular}}
\end{table}

\begin{table}[tpb]
	\caption{\label{Table8} Same as Table~\ref{Table7}, but for the benchmark points of Type-C and -D samples in Table \ref{Table4}, labeled as P3 and P4, respectively. }
	\vspace{0.3cm}
	\resizebox{1.0\textwidth}{!}{
		\begin{tabular}{llll|llll}
			\hline\hline
			\multicolumn{4}{l|}{\texttt{Type-C BP P3}}& \multicolumn{4}{l}{\texttt{Type-D BP P4}}                                                                        \\ \hline
			\multicolumn{1}{l}{$\mu$}     & \multicolumn{1}{r}{773.0 GeV} & \multicolumn{1}{l}{$\text{m}_\text{h}$}         &\multicolumn{1}{r|}{125.6 GeV}    & \multicolumn{1}{l}{$\mu$}     & \multicolumn{1}{r}{751.5 GeV} & \multicolumn{1}{l}{$\text{m}_\text{h}$}         &\multicolumn{1}{r}{125.4 GeV} \\
			\multicolumn{1}{l}{$\text{tan}\beta$}     & \multicolumn{1}{r}{54.1} & \multicolumn{1}{l}{$\text{m}_{\text{A}}$}        &\multicolumn{1}{r|}{9252.5 GeV}     & \multicolumn{1}{l}{$\text{tan}\beta$}     & \multicolumn{1}{r}{58.2} & \multicolumn{1}{l}{$\text{m}_{\text{A}}$}        &\multicolumn{1}{r}{2738.6 GeV}     \\
			\multicolumn{1}{l}{$\text{A}_t$}      & \multicolumn{1}{r}{2653.7 GeV} & \multicolumn{1}{l}{$\text{m}_{\text{H}^{\pm}}$}         &\multicolumn{1}{r|}{9229.6 GeV}     & \multicolumn{1}{l}{$\text{A}_t$}      & \multicolumn{1}{r}{2485.9 GeV} &  \multicolumn{1}{l}{$\text{m}_{\text{H}^{\pm}}$}         &\multicolumn{1}{r}{2766.4 GeV}    \\
			\multicolumn{1}{l}{$\text{M}_1$}        & \multicolumn{1}{r}{360.6 GeV} & \multicolumn{1}{l}{$\text{m}_{\tilde{\nu}_{\mu}}$}        &\multicolumn{1}{r|}{363.9 GeV}     & \multicolumn{1}{l}{$\text{M}_1$}        & \multicolumn{1}{r}{264.8 GeV} & \multicolumn{1}{l}{$\text{m}_{\tilde{\nu}_{\mu}}$}        &\multicolumn{1}{r}{593.6 GeV}     \\
			\multicolumn{1}{l}{$\text{M}_2$} & \multicolumn{1}{r}{1091.0 GeV} & \multicolumn{1}{l}{$\text{m}_{\tilde{\chi}_1^0}$}        &\multicolumn{1}{r|}{360.7 GeV}     & \multicolumn{1}{l}{$\text{M}_2$} & \multicolumn{1}{r}{530.4 GeV} & \multicolumn{1}{l}{$\text{m}_{\tilde{\chi}_1^0}$}        &\multicolumn{1}{r}{264.1 GeV}     \\
			\multicolumn{1}{l}{$\text{M}_{\tilde{\mu}_{\text{L}}}$}        & \multicolumn{1}{r}{113.7 GeV} & \multicolumn{1}{l}{$\text{m}_{\tilde{\chi}_2^0}$}        &\multicolumn{1}{r|}{785.0 GeV}     & \multicolumn{1}{l}{$\text{M}_{\tilde{\mu}_{\text{L}}}$}        & \multicolumn{1}{r}{592.0 GeV}  & \multicolumn{1}{l}{$\text{m}_{\tilde{\chi}_2^0}$}        &\multicolumn{1}{r}{545.0 GeV}     \\
			\multicolumn{1}{l}{$\text{M}_{\tilde{\mu}_{\text{R}}}$}        & \multicolumn{1}{r}{1274.6 GeV} & \multicolumn{1}{l}{$\text{m}_{\tilde{\chi}_3^0}$}        &\multicolumn{1}{r|}{$-793.7$ GeV}     & \multicolumn{1}{l}{$\text{M}_{\tilde{\mu}_{\text{R}}}$}  & \multicolumn{1}{r}{263.0 GeV}  & \multicolumn{1}{l}{$\text{m}_{\tilde{\chi}_3^0}$}        &\multicolumn{1}{r}{$-769.1$ GeV}     \\
			\multicolumn{1}{l}{$a_{\mu}^{\text{SUSY}}$}        & \multicolumn{1}{r}{$1.99\times 10^{-9}$} & \multicolumn{1}{l}{$\text{m}_{\tilde{\chi}_4^0}$}        &\multicolumn{1}{r|}{1136.1 GeV}     & \multicolumn{1}{l}{$a_{\mu}^{\text{SUSY}}$}        & \multicolumn{1}{r}{$2.33 \times 10^{-9}$} &   \multicolumn{1}{l}{$\text{m}_{\tilde{\chi}_4^0}$}        &\multicolumn{1}{r}{782.4 GeV}   \\
			\multicolumn{1}{l}{${\Omega h}^2$}        & \multicolumn{1}{r}{0.13} & \multicolumn{1}{l}{$\text{m}_{\tilde{\chi}_1^{\pm}}$}        &\multicolumn{1}{r|}{784.3 GeV}     & \multicolumn{1}{l}{${\Omega h}^2$}        & \multicolumn{1}{r}{0.14} & \multicolumn{1}{l}{$\text{m}_{\tilde{\chi}_1^{\pm}}$}        &\multicolumn{1}{r}{545.2 GeV}     \\
			\multicolumn{1}{l}{$\sigma_{p}^{\text{SI}}$}       & \multicolumn{1}{r}{$6.78 \times 10^{-47}\text{cm}^{2}$} &  \multicolumn{1}{l}{$\text{m}_{\tilde{\chi}_2^{\pm}}$}        &\multicolumn{1}{r|}{1136.2 GeV}    & \multicolumn{1}{l}{$\sigma_{p}^{\text{SI}}$}       & \multicolumn{1}{r}{$4.64 \times 10^{-47}\text{cm}^{2}$} & \multicolumn{1}{l}{$\text{m}_{\tilde{\chi}_2^{\pm}}$}      &\multicolumn{1}{r}{782.9 GeV}     \\
			\multicolumn{1}{l}{$\sigma_{p}^{\text{SD}}$}       & \multicolumn{1}{r}{$4.54 \times 10^{-43}\text{cm}^{2}$}  &   \multicolumn{1}{l}{$\text{m}_{\tilde{\mu}_{\text{L}}}$}        & \multicolumn{1}{r|}{372.0 GeV}   & \multicolumn{1}{l}{$\sigma_{p}^{\text{SD}}$}       & \multicolumn{1}{r}{$4.16\times 10^{-43}\text{cm}^{2}$} & \multicolumn{1}{l}{$\text{m}_{\tilde{\mu}_{\text{L}}}$}        & \multicolumn{1}{r}{599.2 GeV}     \\
		&	 &  \multicolumn{1}{l}{$\text{m}_{\tilde{\mu}_{\text{R}}}$}        & \multicolumn{1}{r|}{1187.7 GeV}   & &  &  \multicolumn{1}{l}{$\text{m}_{\tilde{\mu}_{\text{R}}}$}        & \multicolumn{1}{r}{268.5 GeV}     \\
\hline
			\multicolumn{2}{l}{$\text{N}_{11},\text{N}_{12},\text{N}_{13},\text{N}_{14}$}           & \multicolumn{2}{r|}{$-0.997,\,~~0.003,-0.071,\,~~0.034$}                 & \multicolumn{2}{l}{$\text{N}_{11},\text{N}_{12},\text{N}_{13},\text{N}_{14}$}           & \multicolumn{2}{r}{$-0.998,\,~~0.006, -0.066,\,~~0.024$}                 \\
			\multicolumn{2}{l}{$\text{N}_{21},\text{N}_{22},\text{N}_{23},\text{N}_{24}$}           & \multicolumn{2}{r|}{$0.074,\,~~0.164,-0.698,\,~~0.693$}                 &               \multicolumn{2}{l}{$\text{N}_{21},\text{N}_{22},\text{N}_{23},\text{N}_{24}$}           & \multicolumn{2}{r}{$-0.023,-0.970,\,~~0.197,-0.140$}                 \\
			\multicolumn{2}{l}{$\text{N}_{31},\text{N}_{32},\text{N}_{33},\text{N}_{34}$}           & \multicolumn{2}{r|}{$0.027,-0.029,-0.706,-0.708$}                 &               \multicolumn{2}{l}{$\text{N}_{31},\text{N}_{32},\text{N}_{33},\text{N}_{34}$}           & \multicolumn{2}{r}{$-0.030,\,~~0.041,\,~~0.705,\,~~0.708$}                 \\
			\multicolumn{2}{l}{$\text{N}_{41},\text{N}_{42},\text{N}_{43},\text{N}_{44}$}           & \multicolumn{2}{r|}{$-0.008,\,~~0.986,\,~~0.096,-0.136$}                 &               \multicolumn{2}{l}{$\text{N}_{41},\text{N}_{42},\text{N}_{43},\text{N}_{44}$}           & \multicolumn{2}{r}{$-0.060,\,~~0.239,\,~~0.678,-0.692$}                                  \\ \hline
			\multicolumn{2}{l}{Annihilations}                     & \multicolumn{2}{l|}{Fractions[\%]}        & \multicolumn{2}{l}{Annihilations}                     & \multicolumn{2}{l}{Fractions[\%]}        \\
			\multicolumn{2}{l}{$\tilde{B}-\tilde{\mu}_L$ Co-annihilation}                  & \multicolumn{2}{l|}{92.6}                 & \multicolumn{2}{l}{$\tilde{B}-\tilde{\mu}_R$ Co-annihilation}                  & \multicolumn{2}{l}{95.2}                 \\
			\hline
			\multicolumn{2}{l}{Decays}                            & \multicolumn{2}{l|}{Branching ratios[\%]} & \multicolumn{2}{l}{Decays}                            & \multicolumn{2}{l}{Branching ratios[\%]} \\
			\multicolumn{2}{l}{$\tilde{\chi}_2^0 \to \tilde{\chi}_1^0h/\tilde{\mu}^{\pm}_{\text{L}} \mu^{\mp} /\tilde{\chi}_1^0Z/ \tilde{\nu}_{\mu}\nu_{\mu} $}& \multicolumn{2}{l|}{70.2/15.1/9.7/5.0}	&\multicolumn{2}{l}{$\tilde{\chi}_2^0 \to \tilde{\chi}_1^0h/ \tilde{\chi}_1^0Z/ \tilde{\mu}^{\pm}_{\text{R}} \mu^{\mp}$}& \multicolumn{2}{l}{78.6/10.7/10.2}                 \\
			\multicolumn{2}{l}{$\tilde{\chi}_3^0 \to \tilde{\chi}_1^0Z/ \tilde{\chi}_1^0h/ \tilde{\mu}^{\pm}_{\text{L}} \mu^{\mp}/ \tilde{\nu}_{\mu}\nu_{\mu}$}& \multicolumn{2}{l|}{86.7/10.8/1.6/0.8}                 & \multicolumn{2}{l}{$\tilde{\chi}_3^0 \to \tilde{\chi}_1^{\pm}W^{\mp}/ \tilde{\chi}_2^0Z /\tilde{\chi}_1^0Z/\tilde{\chi}_1^0h / \tilde{\chi}_2^0h / \tilde{\mu}^{\pm}_{\text{R}} \mu^{\mp}$}                           & \multicolumn{2}{l}{58.2/25.3/12.6/2.8/0.5/0.4}                 \\
			\multicolumn{2}{l}{$\tilde{\chi}_4^0 \to \tilde{\nu}_{\mu}\nu_{\mu}/ \tilde{\mu}^{\pm}_{\text{L}} \mu^{\mp}/ \tilde{\chi}_1^{\pm}W^{\mp}/ \tilde{\chi}_3^0 Z/ \tilde{\chi}_2^0 h/ \tilde{\chi}_2^0 Z/ \tilde{\chi}_3^0 h$}& \multicolumn{2}{l|}{26.8/26.0/24.1/11.3/10.9/0.4/0.3}                 & \multicolumn{2}{l}{$\tilde{\chi}_4^0 \to \tilde{\chi}_1^{\pm}W^{\mp}/ \tilde{\chi}_2^0 h/ \tilde{\chi}_1^0 h / \tilde{\chi}_1^0 Z/ \tilde{\nu}_{\mu}\nu_{\mu} /\tilde{\chi}_2^0 Z/ \tilde{\mu}^{\pm}_{\text{L}} \mu^{\mp}/ \tilde{\mu}^{\pm}_{\text{R}} \mu^{\mp}$}                           & \multicolumn{2}{l}{58.6/23.8/11.2/2.7/1.3/1.0/0.7/0.6}\\
			\multicolumn{2}{l}{$\tilde{\chi}_1^{\pm} \to \tilde{\chi}_1^0W^{\pm}/ \tilde{\nu}_{\mu}\mu^{\pm}/ \tilde{\mu}^{\pm}_{\text{L}}\nu_{\mu} $}                           & \multicolumn{2}{l|}{79.9/13.7/6.0}                 & \multicolumn{2}{l}{$\tilde{\chi}_1^{\pm} \to \tilde{\chi}_1^0 W^{\pm}/ \tilde{\mu}^{\pm}_{\text{R}} \nu_{\mu}$}                           & \multicolumn{2}{l}{92.5/6.9}                 \\
			\multicolumn{2}{l}{$\tilde{\chi}_2^{\pm} \to \tilde{\mu}^{\pm}_{\text{L}} \nu_{\mu}/ \tilde{\nu}_{\mu}\mu^{\pm}/ \tilde{\chi}_2^0W^{\pm} / \tilde{\chi}_1^{\pm}Z/ \tilde{\chi}_3^0W^{\pm}/ \tilde{\chi}_1^{\pm}h$}                     & \multicolumn{2}{l|}{26.3/26.1/12.8/11.9/11.7/11.2}                 & \multicolumn{2}{l}{$\tilde{\chi}_2^{\pm} \to \tilde{\chi}_2^0W^{\pm}/\tilde{\chi}_1^{\pm}Z/ \tilde{\chi}_1^{\pm}h/ \tilde{\chi}_1^0W^{\pm} /\tilde{\mu}^{\pm}_{\text{R}} \nu_{\mu}/\tilde{\nu}_{\mu}\mu^{\pm}$}                     & \multicolumn{2}{l}{31.4/28.4/24.6/13.3/1.3/0.8}                 \\
			\multicolumn{2}{l}{$\tilde{\mu}_{\text{L}}^{\pm} \to \tilde{\chi}_1^{0}\mu^{\pm}$}                              & \multicolumn{2}{l|}{100}                 & \multicolumn{2}{l}{$\tilde{\mu}_{\text{L}}^{\pm} \to \tilde{\chi}_1^0\mu^{\pm}/ \tilde{\chi}_1^{\pm}\nu_{\mu}/ \tilde{\chi}_2^0\mu^{\pm}/ \tilde{\mu}_{\text{R}}^{\pm}h/\tilde{\mu}_{\text{R}}^{\pm}Z$}                              & \multicolumn{2}{l}{68.8/18.5/9.8/1.5/1.4}                 \\
			\multicolumn{2}{l}{$\tilde{\mu}_{\text{R}}^{\pm} \to \tilde{\chi}_1^0\mu^{\pm} /\tilde{\chi}_2^0\mu^{\pm} /\tilde{\nu}_{\mu}W^{\pm} / \tilde{\chi}_1^{\pm}\nu_{\mu} $}                              & \multicolumn{2}{l|}{98.7/0.3/0.3/0.2}                 & \multicolumn{2}{l}{$\tilde{\mu}_{\text{R}}^{\pm} \to \tilde{\chi}_1^0\mu^{\pm}$} &\multicolumn{2}{l}{100}\\
			\multicolumn{2}{l}{$\tilde{\nu}_{\mu} \to \tilde{\chi}_1^{0}\nu_{\mu}$}
			& \multicolumn{2}{l|}{100}                 & \multicolumn{2}{l}{$\tilde{\nu}_{\mu} \to \tilde{\chi}_1^{0}\nu_{\mu}/ \tilde{\chi}_1^{\pm}\mu^{\mp}/ \tilde{\chi}_2^0\nu_{\mu}/ \tilde{\mu}^{\pm}_{\text{R}} W^{\mp}$}                             & \multicolumn{2}{l}{73.0/16.3/7.8/2.8}                 \\
			\hline
			\multicolumn{2}{l}{$R$ value}                           & \multicolumn{2}{l|}{ 0.38, \textbf{SR\_A44} in \texttt{SR-2}}                 & \multicolumn{2}{l}{$R$ value}                           & \multicolumn{2}{l}{0.50, \textbf{SS15} in \texttt{SR-2}}                 \\ \hline\hline
	\end{tabular}}
\end{table}

\begin{table}[tpb]
	\caption{\label{Table9} Benchmark points P5 for Type-E samples in Table \ref{Table4} and P6 located within the blue arc on the $|m_{\tilde{\chi}_1^0}|-m_{\tilde{\mu}_L}$ plane in Fig.~\ref{fig5}. P5 satisfies all the restrictions, while P6 is excluded by the LHC search for SUSY.}
	\vspace{0.3cm}
	\resizebox{1.0\textwidth}{!}{
		\begin{tabular}{llll|llll}
			\hline\hline
			\multicolumn{4}{l|}{\texttt{Type-E BP P5}}& \multicolumn{4}{l}{\texttt{BP P6 for the curved blue area in Fig.~\ref{fig5}}}                                                                        \\ \hline
			\multicolumn{1}{l}{$\mu$}     & \multicolumn{1}{r}{719.2 GeV} & \multicolumn{1}{l}{$\text{m}_\text{h}$}         &\multicolumn{1}{r|}{125.7 GeV}         & \multicolumn{1}{l}{$\mu$}     & \multicolumn{1}{r}{551.2 GeV} & \multicolumn{1}{l}{$\text{m}_\text{h}$}         &\multicolumn{1}{r}{125.2 GeV}     \\
			\multicolumn{1}{l}{$\text{tan}\beta$}     & \multicolumn{1}{r}{38.7} & \multicolumn{1}{l}{$\text{m}_{\text{A}}$}        &\multicolumn{1}{r|}{3235.4 GeV}     & \multicolumn{1}{l}{$\text{tan}\beta$}     & \multicolumn{1}{r}{31.7} & \multicolumn{1}{l}{$\text{m}_{\text{A}}$}        &\multicolumn{1}{r}{6279.6 GeV}     \\
			\multicolumn{1}{l}{$\text{A}_t$}      & \multicolumn{1}{r}{2697.6 GeV} & \multicolumn{1}{l}{$\text{m}_{\text{H}^{\pm}}$}         &\multicolumn{1}{r|}{3253.2 GeV} & \multicolumn{1}{l}{$\text{A}_t$}      & \multicolumn{1}{r}{2398.8 GeV} & \multicolumn{1}{l}{$\text{m}_{\text{H}^{\pm}}$}         &\multicolumn{1}{r}{6284.5 GeV}     \\
			\multicolumn{1}{l}{$\text{M}_1$}        & \multicolumn{1}{r}{277.0 GeV} & \multicolumn{1}{l}{$\text{m}_{\tilde{\nu}_{\mu}}$}        &\multicolumn{1}{r|}{310.2 GeV}     & \multicolumn{1}{l}{$\text{M}_1$}        & \multicolumn{1}{r}{$-304.9$ GeV} & \multicolumn{1}{l}{$\text{m}_{\tilde{\nu}_{\mu}}$}        &\multicolumn{1}{r}{457.4 GeV}     \\
			\multicolumn{1}{l}{$\text{M}_2$} & \multicolumn{1}{r}{1263.2 GeV} &   \multicolumn{1}{l}{$\text{m}_{\tilde{\chi}_1^0}$}        &\multicolumn{1}{r|}{276.0 GeV}   & \multicolumn{1}{l}{$\text{M}_2$} & \multicolumn{1}{r}{317.5 GeV} &     \multicolumn{1}{l}{$\text{m}_{\tilde{\chi}_1^0}$}        &\multicolumn{1}{r}{$-304.1$ GeV} \\
			\multicolumn{1}{l}{$\text{M}_{\tilde{\mu}_{\text{L}}}$}        & \multicolumn{1}{r}{297.4 GeV} & \multicolumn{1}{l}{$\text{m}_{\tilde{\chi}_2^0}$}        &\multicolumn{1}{r|}{731.5 GeV}     & \multicolumn{1}{l}{$\text{M}_{\tilde{\mu}_{\text{L}}}$}        & \multicolumn{1}{r}{409.6 GeV}  &  \multicolumn{1}{l}{$\text{m}_{\tilde{\chi}_2^0}$}        &\multicolumn{1}{r}{327.2 GeV}    \\
			\multicolumn{1}{l}{$\text{M}_{\tilde{\mu}_{\text{R}}}$}        & \multicolumn{1}{r}{288.2 GeV}  & \multicolumn{1}{l}{$\text{m}_{\tilde{\chi}_3^0}$}        &\multicolumn{1}{r|}{$-736.8$ GeV}     & \multicolumn{1}{l}{$\text{M}_{\tilde{\mu}_{\text{R}}}$}        & \multicolumn{1}{r}{755.9 GeV} & \multicolumn{1}{l}{$\text{m}_{\tilde{\chi}_3^0}$}        &\multicolumn{1}{r}{$-570.1$ GeV}     \\
			\multicolumn{1}{l}{$a_{\mu}^{\text{SUSY}}$}        & \multicolumn{1}{r}{$2.11\times 10^{-9}$} & \multicolumn{1}{l}{$\text{m}_{\tilde{\chi}_4^0}$}        &\multicolumn{1}{r|}{1301.3 GeV}     & \multicolumn{1}{l}{$a_{\mu}^{\text{SUSY}}$}        & \multicolumn{1}{r}{$2.03 \times 10^{-9}$} & \multicolumn{1}{l}{$\text{m}_{\tilde{\chi}_4^0}$}        &\multicolumn{1}{r}{578.7 GeV}     \\
			\multicolumn{1}{l}{${\Omega h}^2$}        & \multicolumn{1}{r}{0.13} & \multicolumn{1}{l}{$\text{m}_{\tilde{\chi}_1^{\pm}}$}        &\multicolumn{1}{r|}{730.8 GeV}     & \multicolumn{1}{l}{${\Omega h}^2$}        & \multicolumn{1}{r}{0.10} & \multicolumn{1}{l}{$\text{m}_{\tilde{\chi}_1^{\pm}}$}        &\multicolumn{1}{r}{327.5 GeV}     \\
			\multicolumn{1}{l}{$\sigma_{p}^{\text{SI}}$}       & \multicolumn{1}{r}{$5.44 \times 10^{-47}\text{cm}^{2}$} & \multicolumn{1}{l}{$\text{m}_{\tilde{\chi}_2^{\pm}}$}        &\multicolumn{1}{r|}{1301.5 GeV}     & \multicolumn{1}{l}{$\sigma_{p}^{\text{SI}}$}       & \multicolumn{1}{r}{$1.09 \times 10^{-46}\text{cm}^{2}$} & \multicolumn{1}{l}{$\text{m}_{\tilde{\chi}_2^{\pm}}$}      &\multicolumn{1}{r}{581.0 GeV}     \\
			\multicolumn{1}{l}{$\sigma_{p}^{\text{SD}}$}       & \multicolumn{1}{r}{$5.12 \times 10^{-43}\text{cm}^{2}$} & \multicolumn{1}{l}{$\text{m}_{\tilde{\mu}_{\text{L}}}$}        & \multicolumn{1}{r|}{321.0 GeV}     & \multicolumn{1}{l}{$\sigma_{p}^{\text{SD}}$}       & \multicolumn{1}{r}{$2.04 \times 10^{-42}\text{cm}^{2}$} &       \multicolumn{1}{l}{$\text{m}_{\tilde{\mu}_{\text{L}}}$}        & \multicolumn{1}{r}{464.5 GeV} \\
			& & \multicolumn{1}{l}{$\text{m}_{\tilde{\mu}_{\text{R}}}$}        & \multicolumn{1}{r|}{279.6 GeV}    & & & \multicolumn{1}{l}{$\text{m}_{\tilde{\mu}_{\text{R}}}$}        & \multicolumn{1}{r}{700.3 GeV}      \\ \hline
			\multicolumn{2}{l}{$\text{N}_{11},\text{N}_{12},\text{N}_{13},\text{N}_{14}$}           & \multicolumn{2}{r|}{$-0.997,\,~~0.002,-0.071,\,~~0.028$}                 & \multicolumn{2}{l}{$\text{N}_{11},\text{N}_{12},\text{N}_{13},\text{N}_{14}$}           & \multicolumn{2}{r}{$-0.993,-0.007, -0.106,-0.055$}                 \\
			\multicolumn{2}{l}{$\text{N}_{21},\text{N}_{22},\text{N}_{23},\text{N}_{24}$}           & \multicolumn{2}{r|}{$\,~~0.070,\,~~0.100,-0.703,\,~~0.700$}                 &               \multicolumn{2}{l}{$\text{N}_{21},\text{N}_{22},\text{N}_{23},\text{N}_{24}$}           & \multicolumn{2}{r}{$0.009,\,~~0.971,-0.206,\,~~0.122$}                 \\
			\multicolumn{2}{l}{$\text{N}_{31},\text{N}_{32},\text{N}_{33},\text{N}_{34}$}           & \multicolumn{2}{r|}{$\,~~0.030,-0.027,-0.706,-0.708$}                 &               \multicolumn{2}{l}{$\text{N}_{31},\text{N}_{32},\text{N}_{33},\text{N}_{34}$}           & \multicolumn{2}{r}{$-0.114,\,~~0.060,\,~~0.697,\,~~0.705$}                 \\
			\multicolumn{2}{l}{$\text{N}_{41},\text{N}_{42},\text{N}_{43},\text{N}_{44}$}           & \multicolumn{2}{r|}{$-0.004,\,~~0.995,\,~~0.052,-0.090$}                 &               \multicolumn{2}{l}{$\text{N}_{41},\text{N}_{42},\text{N}_{43},\text{N}_{44}$}           & \multicolumn{2}{r}{$0.036,-0.232,-0.679,\,~~0.696$}                                  \\ \hline
			\multicolumn{2}{l}{Annihilations}                     & \multicolumn{2}{l|}{Fractions[\%]}        & \multicolumn{2}{l}{Annihilations}                     & \multicolumn{2}{l}{Fractions[\%]}        \\
			\multicolumn{2}{l}{$\tilde{B}-\tilde{\mu}_R$ Co-annihilation}                  & \multicolumn{2}{l|}{94.2}                 & \multicolumn{2}{l}{$\tilde{B}-\tilde{W}$ Co-annihilation}                  & \multicolumn{2}{l}{91.4}                 \\
			\hline
			\multicolumn{2}{l}{Decays}                            & \multicolumn{2}{l|}{Branching ratios[\%]} & \multicolumn{2}{l}{Decays}                            & \multicolumn{2}{l}{Branching ratios[\%]} \\
			\multicolumn{2}{l}{$\tilde{\chi}_2^0 \to \tilde{\chi}_1^0h / \tilde{\chi}_1^0Z/ \tilde{\mu}^{\pm}_{\text{L}} \mu^{\mp}/ \tilde{\mu}^{\pm}_{\text{R}} \mu^{\mp} / \tilde{\nu}_{\mu}\nu_{\mu}$}& \multicolumn{2}{l|}{73.1/14.4/7.5/3.6/1.4}   &    \multicolumn{2}{l}{$\tilde{\chi}_2^0 \to \tilde{\chi}_1^0Z^*/\tilde{\chi}_1^0h^*$}& \multicolumn{2}{l}{100}                 \\
			\multicolumn{2}{l}{$\tilde{\chi}_3^0 \to \tilde{\chi}_1^0Z/ \tilde{\chi}_1^0h/  \tilde{\mu}^{\pm}_{\text{R}} \mu^{\mp}/ \tilde{\mu}^{\pm}_{\text{L}} \mu^{\mp}/ \tilde{\nu}_{\mu}\nu_{\mu}$}& \multicolumn{2}{l|}{82.2/14.8/1.1/0.9/0.8}                 & \multicolumn{2}{l}{$\tilde{\chi}_3^0 \to \tilde{\chi}_1^{\pm}W^{\mp}/ \tilde{\chi}_2^0Z / \tilde{\chi}_1^0h/ \tilde{\chi}_2^0h/ \tilde{\chi}_1^0Z$}                           & \multicolumn{2}{l}{62.1/27.0/8.1/1.4/1.1}                 \\
			\multicolumn{2}{l}{$\tilde{\chi}_4^0 \to \tilde{\chi}_1^{\pm}W^{\mp}/ \tilde{\nu}_{\mu}\nu_{\mu}/ \tilde{\mu}^{\pm}_{\text{L}}\mu^{\mp}/ \tilde{\chi}_2^0 h/ \tilde{\chi}_3^0 Z/ \tilde{\chi}_2^0 Z/ \tilde{\chi}_3^0 h/ \tilde{\mu}^{\pm}_{\text{R}}\mu^{\mp}$}& \multicolumn{2}{l|}{25.2/25.1/24.2/11.6/11.5/0.9/0.8/0.4}                 & \multicolumn{2}{l}{$\tilde{\chi}_4^0 \to \tilde{\chi}_1^{\pm}W^{\mp}/  \tilde{\chi}_2^0 h/ \tilde{\chi}_1^0 Z/ \tilde{\chi}_2^0 Z/  \tilde{\chi}_1^0 h/ \tilde{\nu}_{\mu}\nu_{\mu}/ \tilde{\mu}^{\pm}_{\text{L}}\mu^{\mp}$}                           & \multicolumn{2}{l}{63.6/24.0/8.2/2.2/0.8/0.7/0.5}\\
			\multicolumn{2}{l}{$\tilde{\chi}_1^{\pm} \to \tilde{\chi}_1^0 W^{\pm}/ \tilde{\nu}_{\mu}\mu^{\pm}/ \tilde{\mu}^{\pm}_{\text{L}}\nu_{\mu}/ \tilde{\mu}^{\pm}_{\text{R}}\nu_{\mu}$}                           & \multicolumn{2}{l|}{89.7/7.0/1.7/1.1}                 & \multicolumn{2}{l}{$\tilde{\chi}_1^{\pm} \to \tilde{\chi}_1^0(W^{\pm})^*$}                           & \multicolumn{2}{l}{100}                 \\
			\multicolumn{2}{l}{$\tilde{\chi}_2^{\pm} \to \tilde{\nu}_{\mu}\mu^{\pm}/ \tilde{\mu}^{\pm}_{\text{L}}\nu_{\mu}/ \tilde{\chi}_2^0W^{\pm}/\tilde{\chi}_1^{\pm}Z/ \tilde{\chi}_3^0W^{\pm}/\tilde{\chi}_1^{\pm}h  $}                     & \multicolumn{2}{l|}{24.8/24.4/12.8/12.5/12.5/12.4}                 & \multicolumn{2}{l}{$\tilde{\chi}_2^{\pm} \to \tilde{\chi}_2^0W^{\pm}/ \tilde{\chi}_1^{\pm}Z/ \tilde{\chi}_1^{\pm}h/ \tilde{\chi}_1^0W^{\pm}/ \tilde{\mu}^{\pm}_{\text{L}}\nu_{\mu}/ \tilde{\nu}_{\mu}\mu^{\pm}$}                     & \multicolumn{2}{l}{32.3/30.4/25.7/10.3/0.9/0.4}                 \\
			\multicolumn{2}{l}{$\tilde{\mu}_{\text{L}}^{\pm} \to \tilde{\chi}_1^{0}\mu^{\pm}$}                              & \multicolumn{2}{l|}{100}                 & \multicolumn{2}{l}{$\tilde{\mu}_{\text{L}}^{\pm} \to \tilde{\chi}_1^{\pm}\nu_{\mu}/\tilde{\chi}_2^{0}\mu^{\pm}/\tilde{\chi}_1^{0}\mu^{\pm}$}                              & \multicolumn{2}{l}{57.2/29.8/13.0}                 \\
			\multicolumn{2}{l}{$\tilde{\mu}_{\text{R}}^{\pm} \to \tilde{\chi}_1^{0}\mu^{\pm}$}   & \multicolumn{2}{l|}{100} & \multicolumn{2}{l}{$\tilde{\mu}_{\text{R}}^{\pm} \to \tilde{\chi}_1^0\mu^{\pm}/ \tilde{\chi}_3^0\mu^{\pm}$} &\multicolumn{2}{l}{99.5/0.2}\\
			\multicolumn{2}{l}{$\tilde{\nu}_{\mu} \to \tilde{\chi}_1^{0}\nu_{\mu}$}
			& \multicolumn{2}{l|}{100}                 & \multicolumn{2}{l}{$\tilde{\nu}_{\mu} \to \tilde{\chi}_1^{\pm} \mu^{\mp}/ \tilde{\chi}_2^{0}\nu_{\mu}/ \tilde{\chi}_1^{0}\nu_{\mu}$}  & \multicolumn{2}{l}{59.2/28.5/12.3}                 \\
			\hline
			\multicolumn{2}{l}{$R$ value}                           & \multicolumn{2}{l|}{0.34, \textbf{SR\_A44} in \texttt{SR-2}}                 & \multicolumn{2}{l}{$R$ value}                           & \multicolumn{2}{l}{1.22, \textbf{SR\_WZoff\_high\_njd} in \texttt{SR-4}}                 \\ \hline\hline
	\end{tabular}}
\end{table}

\begin{table}[tpb]
\centering
\caption{\label{Table10} Survival mechanisms for the six benchmark points listed in Tables~\ref{Table7}--\ref{Table9}. These mechanisms rely on both the properties of the points and the experimental search strategies. They are acquired by comparing the dominant signals of the productions with relevant experimental analyses. }

\vspace{0.3cm}

	\resizebox{0.99\textwidth}{!}{
\begin{tabular}{|c|c|c|c|c|c|c|}
\hline
 \diagbox{Production}{Survival mechanism }{Points} & P1 & P2 & P3 & P4 & P5 & P6 \\ \hline
 $\bar{\tilde{W}} \tilde{W}$ & I, II~\cite{ATLAS:2021moa}& I, II~\cite{ATLAS:2019lng} & III, IV~\cite{CMS:2018szt} & I, IV~\cite{CMS:2018szt}  & IV~\cite{CMS:2018szt} & I, II~\cite{ATLAS:2021moa} \\ \hline
 $\bar{\tilde{H}} \tilde{H}$ & I, III, IV~\cite{ATLAS:2021moa} & III, IV~\cite{ATLAS:2019lng} & IV~\cite{CMS:2018szt} & III, IV~\cite{CMS:2018szt}  & IV~\cite{CMS:2018szt} & III~\cite{ATLAS:2021moa} \\ \hline
 $\tilde{\mu}_L^* \tilde{\mu}_L$ & I, III, IV~\cite{CMS:2020bfa} & I, II~\cite{ATLAS:2019lng} & I, II~\cite{ATLAS:2019lng}  & IV~\cite{CMS:2020bfa} & I, II~\cite{ATLAS:2019lng}  & III~\cite{CMS:2020bfa} \\ \hline
 $\tilde{\mu}_R^* \tilde{\mu}_R$ & I, IV~\cite{CMS:2020bfa} & IV~\cite{CMS:2020bfa}  & IV~\cite{CMS:2020bfa} & I, II~\cite{ATLAS:2019lng} & I, II~\cite{ATLAS:2019lng}  & IV~\cite{CMS:2020bfa}  \\ \hline
\end{tabular}}
\end{table}

To simplify the discussion of the LHC restrictions, we first classified the blue samples in Fig.~\ref{fig3} by the DM annihilation mechanisms, similar to what we did in Table~\ref{Table5}. Then, we only focused on the samples studied by the Monte Carlo simulations. We show in Table~\ref{Table6} the numbers before and after simulating the LHC restrictions, the SRs that contributed to the largest $R$-values, and their capability to exclude the samples of each mechanism, expressed by the percentage of the total numbers in the second column. The following SRs are involved:
\begin{itemize}
    \item \texttt{SR-1}: Signal regions \textbf{SRG08\_0j\_mll} and \textbf{SRG07\_0j\_mll} defined in Ref.~\cite{CMS:2020bfa}. They come from the LHC search for slepton pair production by the final state containing two opposite-sign-same-flavor (OSSF) leptons and missing transverse momentum.
    \item \texttt{SR-2}: Signal regions \textbf{SR\_A44}, \textbf{SS15}, and \textbf{SR\_A08} in Ref.~\cite{CMS:2017moi,CMS:2018szt}. They arise from the LHC search for electroweakinos with the final state containing missing transverse momentum, no jets, and two same-sign (SS) dileptons for the SS15, and three electrons, or muons that form at least one OSSF pair for the SR\_A44 and SR\_A08.
    \item \texttt{SR-3}: Signal regions \textbf{E−high−mm−30}, \textbf{S−high−mm−05}, \textbf{S−high−mm−10}, and so on proposed in Ref.~\cite{ATLAS:2019lng}. They concentrated on the electroweakinos with compressed mass spectra and investigated the sparticles' production at the LHC by the final state containing two leptons and missing transverse momentum.
    \item \texttt{SR-4}: Signal regions \textbf{SR\_incWZoff\_high\_njc1}, \textbf{SR\_WZoff\_high\_njd}, \textbf{SR\_W-Z\_off\_high\_njc}, and so on defined in Ref.~\cite{ATLAS:2021moa}. They studied the chargino--neutralino associated production at the LHC by the final state containing three leptons and missing transverse momentum, where the chargino and neutralino decayed into off-shell $W$ and $Z$ bosons, respectively.
    \item \texttt{SR-5}: Signal regions \textbf{SR1\_weakino\_3high\_mll\_2}, \textbf{SR2\_stop\_3high\_pt\_1}, and \textbf{SR1\_weakino\_2media\_mll\_2} defined in Ref.~\cite{CMS:2018kag}. They arised from the LHC search for new physics by the signal containing two soft oppositely charged leptons and missing transverse momentum.
\end{itemize}
These analyses revealed how the annihilation mechanisms have been tested at the LHC. In particular, Table~\ref{Table6} shows that different SRs complement each other in doing so, and a single SR never plays a dominant role in this regard. This conclusion depends not only on the intrinsic physics of the MSSM but also on the details of these SRs. It arises from the fact that we included as many experimental analyses as possible to study the LHC restrictions, and each of them usually defined several signal regions. This situation allowed us to make good use of the experimental data to explore the parameter space of the MSSM.

Next, we illuminate how the MSSM manages to survive the LHC restriction. We looked for benchmark points from the Type-A, B, C, D, and E samples specified in Table~\ref{Table4} and labeled them P1, P2, P3, P4, and P5, respectively. We presented their details in Tables~\ref{Table7}, \ref{Table8}, and \ref{Table9}, and provided corresponding
survival mechanisms in Table~\ref{Table10}. These mechanisms were acquired by comparing the dominant signals of the productions with pertinent experimental data. For example, the DM in P1 achieved the measured density by co-annihilating with wino-like particles. The applicable restriction to the wino productions came from the search for the chargino--neutralino associated production, where the chargino and neutralino decayed into off-shell $W$ and $Z$ bosons, respectively. The most robust results came from the analysis in Ref.~\cite{ATLAS:2021moa}, and they served as the guidelines of our judgment. We also provided a benchmark point, P6, located within the blue arc on the $|m_{\tilde{\chi}_1^0}|-m_{\tilde{\mu}_L}$ plane in Fig.~\ref{fig5}, characterized by $ 230~{\rm GeV} \lesssim | m_{\tilde{\chi}_1^0}| \lesssim 350~{\rm GeV}$ and $ 300~{\rm GeV} \lesssim m_{\tilde{\mu}_L} \lesssim 500~{\rm GeV}$. Our simulation indicated that it is excluded by \textbf{SR\_WZoff\_high\_njd} in \texttt{SR-4}, which was designed for the tri-lepton signal from the decays of off-shell $W$ and $Z$ bosons. This point is distinct because both the wino pair production and the $\tilde{\mu}_L$ pair production contribute to this signal, and none of these processes alone can exclude this point. In addition, the program \textsf{SModelS\,2.2.1} is also unable to exclude it since it misses the $\tilde{\mu}_L$ contribution.

At this stage, we clarify the following points about the LHC restrictions:
\begin{itemize}
\item Throughout this study, both the theoretical uncertainties incurred in the simulations and the experimental (systematic and statistic) uncertainties were not included. Although these effects can relax the LHC restrictions, it is expected that much stronger restrictions on the MSSM will be acquired given the advent of high-luminosity LHC in the near future.

\item We did not include the search for charginos and neutralinos by the fully hadronic final states of $W/Z$ and Higgs bosons with $139~{\rm fb}^{-1}$ data~\cite{ATLAS:2021yqv} in this study. This search rejected large SM backgrounds by identifying high-$p_T$ bosons with large-radius jets and jet substructure information. Thus, it was more efficient than the leptonic signal search in Ref.~\cite{ATLAS:2021moa} only when winos were heavier than $600~{\rm GeV}$~\cite{ATLAS:2021yqv}. This conclusion holds in simplified models of SUSY but is never applied to this research. The reason is that the DM must co-annihilate with $\tilde{\mu}_L$ or $\tilde{\mu}_R$ to achieve the measured density for $M_2 \gtrsim 600~{\rm GeV}$ (see the results in Fig.~\ref{fig2}), and the wino-like particles may decay into the slepton first to enhance their leptonic signal. As a result, it is the leptonic signals that are more powerful in excluding the SUSY points. This conclusion is the same as that of the General Next-to-Minimal Supersymmetric Standard Model (GNMSSM) studied in Ref.~\cite{Cao:2022ovk}.

    The latest version of \textsf{SModelS}, namely \textsf{SModelS\,2.2.1}~\cite{Alguero:2021dig}, have implemented the cut efficiencies of the hadronic analysis and relevant signal topologies in its database. We utilized this code to restrict the green samples in Fig.~\ref{fig1}. We did not find that it had exclusion capabilities.

\item In some high-energy SUSY-breaking theories, $\tilde{\tau}$ may be the NLSP due to its larger Yukawa coupling than those of the first- and second-generation sleptons. In this case, heavy sparticles may decay into $\tilde{\tau}$ to change the $e/\mu$ signals of this study, and the LHC restrictions may be relaxed~\cite{Hagiwara:2017lse}. We will discuss such a possibility in our future work.
\end{itemize}

\subsection{Related issues}

We stress the following issues related to this research:

\begin{itemize}
\item Throughout this study, we assumed that $\tilde{\chi}_1^0$ was fully responsible for the measured density. This assumption determined that the DM is bino-dominated and approximately degenerate with winos or sleptons in mass. This has profound implications on the phenomenology of the MSSM. Relaxing this assumption usually complicates this kind of research and makes the obtained conclusions untenable. The study in Ref.~\cite{Chakraborti:2021kkr} illustrated this point, where the authors replaced the requirement on the DM relic density in this work, i.e., $0.096 \leq \Omega h^2 \leq 0.144$, with $\Omega {h^2} \leq 0.120$, and studied the restrictions of various experiments, including the DM direct detection experiments and the LHC searches for SUSY, on interpreting the muon g-2 anomaly with the MSSM. They concluded that the preferred DM candidate might be the higgsino-, wino-, or bino-dominated neutralino. In either case, the restrictions became significantly weak compared with the results of this study, and consequently, broader parameter spaces could account for the anomaly. Let's take the wino-dominated DM as an example. It was found that $m_{\tilde{\chi}_1^0}$ and $\mu$ might be as low as $100~{\rm GeV}$ and $300~{\rm GeV}$, respectively, without contradicting the restrictions (see Table 2 of Ref.~\cite{Chakraborti:2021kkr}). In this case, the relic density was around $10^{-4}$, and the SI DM-nucleon scattering cross-section was below $2 \times 10^{-47}~{\rm cm^2}$. The wino-like $\tilde{\chi}_1^\pm$ was invisible at the LHC due to its approximate degeneracy with $\tilde{\chi}_1^0$ in mass. The rates of the di- and tri-lepton signals from the higgsino-like electroweakino pair productions were suppressed by the multiple decay possibilities of the produced particles and thus kept consistent with the results of the LHC search for SUSY.

\item As pointed out by the recent lattice simulation of the BMW collaboration on the hadronic
vacuum polarization (HVP) contribution to $a_\mu$~\cite{Borsanyi:2020mff}, the muon g-2 anomaly might arise from the uncertainties in calculating the hadronic contribution to the moment. If this speculation is corroborated, $a_\mu^{\rm SUSY}$ should be much smaller than its currently favored size, and any of the electroweakinos and $\tilde{\mu}_{L/R}$ are not necessarily light. In this case, the LHC restrictions will be relaxed significantly. For example, we recently updated the results of Ref.~\cite{Cao:2021ljw}, which only studied the DM physics in GNMSSM, by including the recent LZ restrictions. We found that the analyses in Table~\ref{Table1} only excluded about $4\%$ of the remaining samples in Fig. 2 of Ref.~\cite{Cao:2021ljw}.

\item It has been argued that the explanations of the anomaly with the MSSM will be tested at future colliders, given that they predict some moderately light sparticles, such as $m_{\tilde{\chi}_1^0}$, $m_{\rm NLSP}$, and $m_{\rm NNLSP}$. For example, the authors of Ref.~\cite{Chakraborti:2021squ} compared the capabilities of different colliders to scrutinize the explanations and presented their results in Fig. 4 for the $\tilde{B}-\tilde{W}$ co-annihilation case. They found that although only parts of the preferred parameter space could be covered in the high-luminosity LHC, exhaustive coverage of the parameter space was possible at a high-energy $e^+ e^-$ collider with $\sqrt{s} \gtrsim 1~{\rm TeV}$, such as ILC with $\sqrt{s} = 1~{\rm TeV}$~\cite{ILC:2013jhg} and CLIC with $\sqrt{s} = 1~{\rm TeV}$~\cite{CLICDetector:2013tfe,CLICdp:2018cto}. We realize that this conclusion is conditionally valid. One exception corresponds to the case that the central value of the anomaly is significantly reduced, introduced in the last item, so that  multi-TeV scale supersymmetric theories are still capable of accounting for the anomaly. Another exception is the tremendously massive higgsino scenario discussed in Ref.~\cite{Gu:2021mjd}, where all sparticles are heavier than $1~{\rm TeV}$, and in particular, the higgsino mass is typically several tens TeV. This situation is remarkable since it implies that the fine-tuning criteria will no longer serve as a valuable guideline to build new physics theories, assuming that $a_\mu^{\rm SUSY} \sim 2.0 \times 10^{-9}$ is always needed to account for the anomaly, and simultaneously no SUSY signals are detected in the future.
\end{itemize}

\section{\label{conclusion-section}Summary}

Inspired by the rapid progress of particle physics experiments in recent years, we studied their impacts on the MSSM and how the theory kept consistent with them.
We are particularly interested in the recent measurement of the muon g-2 at Fermilab, the LHC search for SUSY, and the DM direct detection by the LZ experiment since they are sensitive to different parameters and complement each other to provide valuable information of the MSSM. In surveying the status of the MSSM, we utilized the MultiNest algorithm to comprehensively scan its parameter space. We adopted the muon g-2 observable to guide the scans and included the restrictions from the LHC Higgs data, DM experiments, B-physics measurements, and vacuum stability. We also examined the samples acquired from the scans by the restrictions from the LHC search for SUSY and the latest LZ experiment. The main conclusions of this research are as follows:
\begin{itemize}
\item The bino-dominated DM achieves the measured relic density by co-annihilating with wino-like electroweakinos, $\tilde{\mu}_L$, or $\tilde{\mu}_R$ if one intends the theory to explain the muon g-2 anomaly.
\item Given the measured DM density, the LZ experiment alone has required $\mu \gtrsim 380~{\rm GeV}$ for $M_1 < 0 $ and $\mu \gtrsim 600~{\rm GeV}$ for $M_1 > 100~{\rm GeV}$. If the restrictions from the muon g-2 anomaly and the LHC experiment are also included, the lower bounds become about $500~{\rm GeV}$ and $630~{\rm GeV}$, respectively, indicating that the theory needs a tuning of ${\cal{O}}(1\%)$ to predict $Z$-boson mass~\cite{Baer:2012uy}. Compared with the restriction from the PandaX-4T experiment on the SI scattering cross section, these bounds are improved by about $100~{\rm GeV}$. This situation makes the $Z$- and $h$-mediated resonant annihilations even more unnatural so that they are usually missed in the scans by the \textsf{MultiNest} algorithm. The fundamental reason of such phenomena is the correlation between the DM physics and the electroweak symmetry breaking in the MSSM.

\item On the premise of explaining the muon g-2 anomaly at the $2 \sigma$ level, the LHC restrictions have set bounds on the sparticle mass spectra: $ m_{\tilde{\chi}_1^0} \gtrsim 210~{\rm GeV}$, $m_{\tilde{\chi}_2^0}, m_{\tilde{\chi}_1^\pm} \gtrsim 235~{\rm GeV}$, $m_{\tilde{\chi}_3^0} \gtrsim 515~{\rm GeV}$, $m_{\tilde{\chi}_4^0} \gtrsim 525~{\rm GeV}$, $m_{\tilde{\chi}_2^\pm} \gtrsim 530~{\rm GeV}$, $m_{\tilde{\nu}_\mu} \gtrsim 235~{\rm GeV}$, $ m_{\tilde{\mu}_1} \gtrsim 215~{\rm GeV}$, and $m_{\tilde{\mu}_2} \gtrsim 250~{\rm GeV}$, where $\tilde{\chi}_{2}^0$ and $\tilde{\chi}_1^\pm$ are wino-dominated when they are lighter than about $500~{\rm GeV}$ and $m_{\tilde{\mu}_1}$ may be either $m_{\tilde{\mu}_L}$ or $m_{\tilde{\mu}_R}$. These bounds should be regarded as rough estimates, instead of accurate values, since the samples studied in this research are far from sufficient given the broad parameter space of the MSSM. In addition, these bounds are far beyond the reach of the LEP experiments in searching for SUSY and have not been acquired before.

    The results can be interpreted as follows: if $\tilde{\chi}_1^0$ is lighter, more missing transverse energy will be emitted in the sparticle production processes at the LHC, which can improve the sensitivities of the experimental analyses; while if the sparticles other than $\tilde{\chi}_1^0$ are lighter, they will be more copiously produced at the LHC to increase the events containing multiple leptons.

\item We illuminate how some parameter spaces of the MSSM have been tested at the LHC in Table~\ref{Table6}. We also list five scenarios that are consistent with the LHC restrictions in Table~\ref{Table4} and explain why they can do so by presenting benchmark points in Tables~\ref{Table7}--\ref{Table9} and their survival mechanisms in Table~\ref{Table10}.
\end{itemize}

This work extends the previous studies of the muon g-2 anomaly in the MSSM, in particular those of Refs. \cite{Chakraborti:2020vjp} and \cite{Baum:2021qzx},
by utilizing more sophisticated research strategies and surveying the LHC restrictions comprehensively. As a result, the conclusions acquired in this research are more robust than those in previous works. They exhibit the most essential characteristics of the MSSM.

{\bf Note added}: Recently, the E989 experiment at Fermilab updated its measurement of muon g-2~\cite{Muong-2:2023cdq}. The new world average of $a_\mu^{\rm Exp}$ showed a $5.1 \sigma$ discrepancy from the SM prediction acquired by the Muon (g-2) Theory Initiative in 2020~\cite{Aoyama:2020ynm}, which used dispersive techniques to extract the leading-order HVP contribution from $e^+ e^- \to$ hadrons data. The difference is now given by~\cite{Muong-2:2023cdq}
\begin{eqnarray}
\Delta a_{\mu} \equiv a^{\rm Exp}_\mu - a^{\rm SM}_{\mu} = (24.9 \pm 4.8) \times 10^{-10}.
\end{eqnarray}
Compared with the previous result in 2021~\cite{Abi:2021gix}, the central value of $\Delta a_{\mu}$ changes slightly, while its uncertainty is significantly reduced. We required the MSSM to explain the updated discrepancy at the $2\sigma$ and $3\sigma$ levels, respectively, and repeated the analyses of this work. We found the main conclusions of this study unchanged.

\section{Acknowledgement}
We sincerely thank Prof. Junjie Cao for numerous helpful discussions and his great efforts to improve the manuscript.
This work is supported by the National Natural Science Foundation of China (NNSFC) under grant Nos. 12075076 and 11905044.

\bibliographystyle{CitationStyle}
\bibliography{MSSM}

\end{document}